\documentclass[a4paper,onecolumn,11pt,accepted=2026-10-12]{quantumarticle}
\pdfoutput=1
\usepackage[utf8]{inputenc}
\usepackage[english]{babel}
\usepackage[T1]{fontenc}
\usepackage{amsmath}
\usepackage{hyperref}

\usepackage{tikz}
\usepackage{lipsum}
\usepackage{graphicx} 
\usepackage{subcaption}
\usepackage{amssymb}
\usepackage{amsthm}
\usepackage{xcolor}
\usepackage{comment}
\usepackage{authblk}
\usepackage[sorting=none, maxbibnames=99, minbibnames=99]{biblatex}
\usepackage{fullpage}
\newcommand{\bra}[1]{\langle#1| }
\newcommand{\ket}[1]{|#1\rangle }

\newcommand{\h}{\mathfrak{h}}

\newcommand{\Tr}{\text{Tr}}

\begin{document}

\title{Failure of the mean-field Hartree approximation for a bosonic many-body system with non-Hermitian Hamiltonian}

\author{Matias Gabriel Ginzburg}
\affiliation{Scuola Internazionale Superiore di Studi Avanzati, Via Bonomea, 265, 34136 Trieste TS, Italy }
\orcid{0000-0001-8834-1202}
\author{Simone Rademacher}
\affiliation{Department of Mathematics, LMU Munich, Theresienstrasse 39, 80333 Munich, Germany}
\orcid{0000-0001-5059-4466}
\author{Giacomo De Palma}
\affiliation{University of Bologna, Department of Mathematics, Piazza di Porta San Donato 5, 40126 Bologna BO, Italy}
\orcid{0000-0002-5064-8695}

\maketitle

\begin{abstract}
Mean-field Hartree theory is a central tool for reducing many-body dynamics to an effective one-particle evolution. This approximation has been employed when the Hamiltonian governing the many-body dynamics is not Hermitian. Indeed, non-Hermitian Hamiltonians model particle gain/loss or the evolution of open quantum systems between consecutive quantum jumps. Furthermore, the validity of the Hartree approximation for generic non-Hermitian Hamiltonians lies at the basis of a quantum algorithm for nonlinear differential equations. In this work, we show that this approximation can fail.

First, we consider the most general bosonic mean-field Hamiltonian with two-body interactions and we prove that, for non-Hermitian Hamiltonian, the Hartree equation does not capture the correct second time derivative of the one-particle marginal state at $t=0$ in the thermodynamic limit.
Then, we analytically solve a model of $N$ bosonic qubits with a purely anti-Hermitian Hamiltonian, compute the limit for $N\to\infty$ of the one-particle marginal state and show that it does not agree with the solution of the non-Hermitian Hartree equation. We further show that there exists an initial condition such that the exact one-particle marginal state undergoes a finite-time transition to a mixed state, a phenomenon that is completely absent in the case of Hermitian Hamiltonians.
Finally, we simulate a non-Hermitian Hamiltonian with real spectrum and show that, also in this case, there are initial conditions for which the Hartree equation does not reproduce the correct one-particle marginal state in the limit $N\to\infty$.
 
Our findings challenge the validity of the mean-field Hartree approximation for non-Hermitian Hamiltonians, and call for additional conditions for the validity of the mean-field regime to model the dynamics of particle gain and loss and the open-system dynamics in bosonic many-body systems.
\end{abstract}

\section{Introduction}

Mean-field theory provides an effective description of bosonic many-body systems with a large number of particles by replacing the full many-body dynamics with a single nonlinear evolution equation for the wavefunction of a one-particle. In this approximation, each particle interacts with an average field generated by the collective state of the other particles. The heuristic of this description is based on the assumption that the entanglement of each individual particle with the rest of the system is suppressed when the number of particles is very large, and therefore the one-particle marginal states remain approximately pure.

In the standard setting of Hermitian Hamiltonians, such a mean-field limit is by now well understood and leads to the Hartree or Gross--Pitaevskii equations \cite{knowles2010mean,BenedikterPortaSchlein2016}. Early rigorous results go back to Hepp \cite{hepp1974classical}. Spohn’s analysis via the BBGKY hierarchy \cite{spohn1981vlasov} provided a general framework for bounded interactions, later extended to Coulomb potentials in \cite{ErdosYau2001}. Subsequent works established quantitative rates of convergence \cite{rodnianski2009quantum, chen2011rate}.
A natural question is whether similarly general results hold for non-Hermitian Hamiltonians. Such models arise effectively in atomic, optical, condensed-matter, and photonic platforms \cite{Aspelmeyer14Cavity,rogers2014hybrid,Rotter2009,ElGanainy2018,BergholtzBudichKunst2021,AshidaGongUeda2020}, where they describe mechanisms of particle gain and loss. Particle losses in many-body systems are often modeled by Markovian master equations of Gorini--Kossakowski--Lindblad--Sudarshan form \cite{GoriniKossakowskiSudarshan1976,Lindblad1976}. Conditioning such dynamics on the absence of quantum jumps leads to an effective \emph{non-unitary} evolution generated by a non-Hermitian Hamiltonian; this quantum-trajectory viewpoint underlies standard methods in quantum optics and atomic, molecular and optical many-body physics \cite{DalibardCastinMolmer1992,Carmichael1993,PlenioKnight1998,Daley2014}. Furthermore, in cold-atom settings, controlled dissipation and losses are not merely imperfections but resources, enabling the preparation of correlated states \cite{Syassen2008,Barontini2013}. 
Mean-field approximations for non-Hermitian systems are used in the physics literature \cite{Graefe_2008, Non_Linear_Transport_Pual_07,Sceli2022}, but no rigorous justification is known yet. Related progress exists for open quantum systems governed by Lindblad dynamics in the thermodynamic limit with clustering  states \cite{fiorelli2023mean, benatti2015dissipative}. However, these results do not directly address the validity of mean-field approximations for general non-Hermitian Hamiltonians, for which no systematic theoretical understanding is available.

These developments motivate a basic question: Does the Hartree approximation for the one-particle marginal state remain valid for generic bosonic non-Hermitian many-body Hamiltonians?

The physics literature, starting from the non-unitary $N$-body Schr\"odinger evolution with a non-Hermitian mean-field Hamiltonian, considers the normalized $k$-particle marginals $\hat{\rho}^{(k)}=\rho^{(k)}/\Tr\rho^{(k)}$, whose dynamics is already nonlinear because of the normalization. Closing the resulting (normalized) BBGKY hierarchy by the usual factorization ansatz for which the few-body marginal states are factorized, \emph{i.e.}, $\hat{\rho}^{(k)}\approx (\hat{\rho}^{(1)})^{\otimes k}$ yields a trace-preserving nonlinear evolution equation for $\hat{\rho}^{(1)}$.
As it is the case for Hermitian Hamiltonians, such an equation preserves the set of pure states, and can therefore be turned into a nonlinear evolution equation for the one-particle wavefunction \cite{Graefe_2008,GraefeKorschNiederle2010,GraefeHoeningKorsch2010,Non_Linear_Transport_Pual_07,BrodyGraefe2012}.

This extension of the Hartree equation to non-Hermitian Hamiltonians has also recently entered the field of quantum computing through a proposal that exploits mean-field embeddings to emulate nonlinear dynamics. In particular, Lloyd \emph{et al.} \cite{NonLinearDiffEq_Lloyd} propose a quantum algorithm for nonlinear differential equations that solves a target nonlinear evolution by simulating a linear evolution on many copies of a system, chosen so that the Hartree reduction of the linear many-copy dynamics reproduces the desired nonlinearity. A key step in that approach is the assumption that the Hartree approximation for the one-particle marginal state is valid for \emph{generic} non-Hermitian many-body Hamiltonians.

The main message of the present work is that, even for extremely simple two-body non-Hermitian interactions, the non-Hermitian Hartree equation can fail to capture the one-particle dynamics.

First, we consider the most general non-Hermitian bosonic mean-field Hamiltonian with two-body interactions, and we compare at $t=0$ the first and the second time derivatives of the solution of the Hartree equation and of the one-particle marginal state in the limit $N\to\infty$.
We find that, while the Hartree equation is always able to correctly capture the first time derivative, it fails to predict the second time derivative if the Hamiltonian is not Hermitian and the interaction terms are nonzero.

The breakdown of the mean-field Hartree approximation is due to the fact that the evolution equation of the one-particle marginal state is affected by the anti-Hermitian terms of the Hamiltonian that do not act on the considered particle.
Indeed, while the Hermitian terms act on the many-body density operator through a commutator, which is canceled out when the particles on which they act are traced out, the anti-Hermitian terms act through an anti-commutator, which does not vanish after taking the partial trace.
Such anti-Hermitian terms collectively generate a contribution that has a prefactor that scales as $O(N)$ and can therefore amplify the deviations of the few-particle marginal states from a product state.

Then, we provide an explicit, exactly solvable counterexample: a bosonic system of $N$ qubits with two-particle interactions generated by a \emph{purely anti-Hermitian} Hamiltonian. For any factorized initial wavefunction in the completely symmetric subspace, we compute the exact normalized one-particle marginal state $\hat{\rho}^{(1)}_N(t)$ and compare its limit for $N\to\infty$ with the solution of the non-Hermitian Hartree equation. Our analytical large-$N$ analysis and supporting numerics show:
\begin{itemize}
\item \textbf{Persistent failure of the Hartree approximation:} For almost any initial condition, the limit for $N\to\infty$ of the one-particle marginal state is pure but different from the solution of the non-Hermitian Hartree equation. Indeed, such a limit is governed by an effective evolution equation that depends explicitly on time and agrees with the Hartree equation only at $t=0$.
\item \textbf{Finite-time mixedness transition:} There exists an initial condition for which the limit of $N\to\infty$ of the one-particle marginal state undergoes a sharp transition from a pure state to a mixed state at a finite critical time, a phenomenon which is completely absent for Hermitian Hamiltonians and is indicative of correlations that survive the mean-field limit.
\end{itemize}

We further consider an example of non-Hermitian mean-field bosonic Hamiltonian with two-body interactions and real spectrum, and we numerically find that also in this case there are initial product states for which the one-particle marginal state of the time-evolved many-body state does not converge to the solution of the non-Hermitian Hartree equation in the limit $N\to\infty$.

These findings have two immediate implications.

First, they place concrete limits on the scope of mean-field embeddings for quantum computation. In the setting of \cite{NonLinearDiffEq_Lloyd}, the correctness of the nonlinear solver hinges on whether the many-copy linear dynamics admits the claimed Hartree approximation for the one-particle marginal state. Our counterexample shows that for non-Hermitian Hamiltonians this approximation may \emph{not} hold: even for two-body exactly solvable interactions, the normalized one-particle marginal state may have a different limit for $N\to\infty$ than the non-Hermitian Hartree evolution. Consequently, additional structural assumptions or a case-by-case validation of the Hartree approximation appear necessary before such embeddings can be used as generic algorithmic primitives.

Second, the results are relevant to many-body modeling with losses. Non-Hermitian Hamiltonians are routinely employed as effective descriptions of particle loss and monitoring, either as stand-alone models in non-Hermitian many-body physics \cite{AshidaGongUeda2020,BergholtzBudichKunst2021} or as conditional no-jump generators within the quantum-trajectory formulation of Lindblad dynamics \cite{DalibardCastinMolmer1992,PlenioKnight1998,Daley2014,ClusterBiella}. Mean-field approximations are then often invoked to arrive at complex Hartree or Gross--Pitaevskii-type equations. Our counterexample highlights that non-unitarity can amplify correlations in a way that prevents the usual mean-field closure. This suggests that modeling particle-loss dynamics in bosonic many-body systems by a non-Hermitian Hartree equation may require extra justification, rather than being taken as a generic consequence of large $N$.

The structure of this article is the following. In \autoref{sec:Non-Hermitian_mean_field}, we introduce the notation and the framework for analyzing the mean-field limit in the non-Hermitian setting, and how the Hartree equation is derived using heuristic methods. We discuss new terms that appear in the evolution equation of the one-particle reduced density matrix that are not captured by the Hartree evolution, and in \autoref{sec:derivatives} we show how these terms modify the second time derivative.

In \autoref{sec:ZZ_hamiltonian}, we construct an analytically solvable counterexample, showing that the heuristic extension of the Hermitian theory fails. The \autoref{sec:Large_N_limit} is dedicated to analytical results in the limit of large $N$ and \autoref{sec:Numerical_Analysis} presents numerical simulations supporting and extending the analytical results. Lastly, in \autoref{sec:ZZ_RealSpectrum} we construct another example, where the Hamiltonian is not Hermitian but it has real spectrum. Numerical simulations show that also in this case the heuristic extension of the Hermitian theory fails.
We conclude in \autoref{sec:concl}.

\section{The Hartree equation for non-Hermitian Hamiltonians}
\label{sec:Non-Hermitian_mean_field}

Let $\h$ be the Hilbert space of a single particle, which we assume to have finite dimension, and let $\mathrm{Sym}^N\h$ be its $N$-fold symmetrized tensor product. Mean-field Hamiltonians are constructed by a one-particle operator $A^{(1)}$ acting on $\h$ and a two-particle operator $A^{(2)}$ acting on $\h^{\otimes2}$ and invariant with respect to the swap of the two particles. The $N$-particle mean-field Hamiltonian $A^{(N)}$ is
\begin{equation}\label{def:mean_field_operator}
    A^{(N)} = \sum_{i=1}^N A^{(1)}_i +\frac{1}{N-1}\sum_{1\leq i<j\leq N} A^{(2)}_{ij}, 
\end{equation}
where sub-indexes indicate on which particles the one-particle and the two-particle operators act.
We stress that $A^{(N)}$ is invariant with respect to arbitrary permutations of the particles, and therefore leaves invariant the completely symmetric subspace $\mathrm{Sym}^N\h$.

Let us consider the non-Hermitian Schr\"odinger equation
\begin{equation}\label{eq:Many-body_Schrodinger_Psi}
\begin{cases}
     i \partial_t |\Psi^{(N)}\rangle = A^{(N)} |\Psi^{(N)}\rangle \\
     |\Psi^{(N)}(0)\rangle = |\varphi\rangle^{\otimes N}
\end{cases}
\end{equation}
for the $N$-particle wavefunction $|\Psi^{(N)}\rangle\in\h^{\otimes N}$, where the initial condition is a factorized wavefunction where each particle has wavefunction $|\varphi\rangle\in\h$.
Without loss of generality, we assume that $\langle\varphi|\varphi\rangle = 1$.
Equation \eqref{eq:Many-body_Schrodinger_Psi} can also be written in terms of the unnormalized $N$-particle state $\rho^{(N)}:= \ket{\Psi^{(N)}}\bra{\Psi^{(N)}}$:
\begin{equation}\label{eq:Many-body_Schrodinger}
\begin{cases}
    i \partial_t \rho^{(N)} = A^{(N)} \rho^{(N)} - \rho^{(N)} {A^{(N)}}^\dagger \\
    \rho^{(N)}(0) = \ket{\varphi}\bra{\varphi}^{\otimes N}.
\end{cases}
\end{equation}
For any $k=1,\,\ldots,\,N-1$, the unnormalized $k$-particle marginal state is defined by
\begin{equation}\label{def:marginal_density_matrix}
\rho^{(k)} = \Tr_{k+1\dots N}\rho^{(N)}\,.
\end{equation}
This definition implies the normalization
\begin{equation}
    \Tr\rho^{(k)} = \Tr\rho^{(N)} = \langle\Psi^{(N)}|\Psi^{(N)}\rangle \quad \forall \;k=1,\,\ldots,\,N-1\,,
\end{equation} 
where the norm of $|\Psi^{(N)}\rangle$ can change in time because the evolution is not unitary. The evolution equation for the unnormalized one-particle marginal state is computed by taking the partial trace of equation \eqref{eq:Many-body_Schrodinger}:
\begin{equation}\label{eq:partial_t_rho_exact}
\begin{split}
    i\partial_t \rho^{(1)} =& A^{(1)}\rho^{(1)} - \rho^{(1)}{A^{(1)}}^\dagger + \Tr_{2}\left(A^{(2)} \rho^{(2)} - \rho^{(2)} {A^{(2)}}^\dagger\right) \\
    +& \left(N-1\right) \Tr_{2}\left(\left(\mathbb{I}\otimes\left(A^{(1)}-{A^{(1)}}^\dagger\right)\right) \rho^{(2)}\right) + \frac{N-2}{2} \Tr_{23}\left(\left(\mathbb{I}\otimes\left(A^{(2)}-{A^{(2)}}^\dagger\right)\right) \rho^{(3)}\right).
\end{split}
\end{equation}
For any $k=1,\,\ldots,\,N$, let
\begin{equation}
    \hat{\rho}^{(k)} = \frac{\rho^{(k)}}{\Tr\rho^{(k)}}
\end{equation}
be the normalized $k$-particle marginal state, which will simply be called $k$-particle marginal state.
The evolution equation for the one-particle marginal state is 
\begin{equation}\label{eq:partial_t_rho_exact_normalized}
\begin{split}
    i \partial_t \hat{\rho}^{(1)} =& A^{(1)}\hat{\rho}^{(1)} - \hat{\rho}^{(1)}{A^{(1)}}^\dagger - \Tr\left(\left(A^{(1)}-{A^{(1)}}^\dagger\right)\hat{\rho}^{(1)}\right) \hat{\rho}^{(1)} \\
    &+ \Tr_{2}\left(A^{(2)} \hat{\rho}^{(2)} - \hat{\rho}^{(2)} {A^{(2)}}^\dagger\right) - \Tr\left(\left(A^{(2)}-{A^{(2)}}^\dagger\right)\hat{\rho}^{(2)}\right) \hat{\rho}^{(1)} \\
    &+\left(N-1\right) \Tr_{2}\left(\left(\mathbb{I}\otimes\left(A^{(1)}-{A^{(1)}}^\dagger\right) \right) (\hat{\rho}^{(2)} - \hat{\rho}^{(1)}{}^{\otimes 2} )\right) \\
    &+ \frac{N-2}{2} \Tr_{23}\left(\left(\mathbb{I}\otimes \left(A^{(2)}-{A^{(2)}}^\dagger\right) \right) \left(\hat{\rho}^{(3)} - \hat{\rho}^{(1)} \otimes \hat{\rho}^{(2)}\right)  \right).
\end{split}
\end{equation}
The first two lines contain the usual terms in the  Hartree equation, plus terms that control the norm in the non-Hermitian case.  The main novelty in the non-Hermitian case is in the last two lines, where the terms are multiplied by a factor $N$ and vanish in the Hermitian case. To obtain a closed evolution equation for $\hat{\rho}^{(1)}$ from equation \eqref{eq:partial_t_rho_exact_normalized}, the heuristic approach that works in the Hermitian case is to postulate that the two-particle marginal state $\hat{\rho}^{(2)}$ approximately remains a product state: $\hat{\rho}^{(2)} \simeq \hat{\rho}^{(1)}{}^{\otimes 2}$.
In the non-Hermitian case, the time derivative of $\hat{\rho}^{(1)}$ depends also on the three-particle marginal state $\hat{\rho}^{(3)}$.
Therefore, we will obtain the nonlinear Schr\"odinger equation by postulating that also the three-particle state remains a product state: $\hat{\rho}^{(3)} \simeq \hat{\rho}^{(1)}{}^{\otimes 3}$. Assuming that all these postulates hold true and replacing $\rho^{(k)}$ by $\gamma^{\otimes k}$ in equation \eqref{eq:partial_t_rho_exact_normalized} for $k=1,\,2,\,3$ gives the following non-Hermitian Hartree equation:
\begin{equation}\label{eq:HF_non_Hermitian_normalized}
\begin{split}
    i \partial_t \gamma &= A^{(1)}\gamma - \gamma{A^{(1)}}^\dagger - \Tr\left(\left(A^{(1)}-{A^{(1)}}^\dagger\right)\gamma\right) \gamma \\
    &\phantom{=} + \Tr_{2}\left( A^{(2)} \left(\mathbb{I} \otimes \gamma\right)\right) \gamma - \gamma\,\Tr_2\left( \left(\mathbb{I} \otimes \gamma\right) {A^{(2)}}^\dagger\right) - \Tr\left( \left(A^{(2)}-{A^{(2)}}^\dagger\right) \gamma{}^{\otimes 2}\right) \gamma\,,
\end{split}
\end{equation}
with initial condition $\gamma(0) = |\varphi\rangle\langle\varphi|$.
We claim that $\gamma$ will be pure and normalized at any time.
Indeed, $\gamma=\ket{\phi}\bra{\phi}$ satisfies \eqref{eq:HF_non_Hermitian_normalized} if the Hartree wavefunction $|\phi\rangle\in\h$ satisfies the non-Hermitian nonlinear Schr\"odinger equation
\begin{equation}\label{eq:HF_non_Hermitian_vector}
    i \partial_t |\phi\rangle = \left(A^{(1)} - \langle\phi| \frac{A^{(1)}-A^{(1)\dagger}}{2} |\phi \rangle + \langle\phi|A^{(2)}|\phi\rangle  - \langle \phi^{\otimes2}|\frac{A^{(2)}-A^{(2)\dagger}}{2} |\phi^{\otimes 2} \rangle \right) |\phi\rangle
\end{equation}
with initial condition $|\phi(0)\rangle = |\varphi\rangle$.
Furthermore, the solution of \eqref{eq:HF_non_Hermitian_vector} has unit norm at any time.

We will analyze two quantities related to the Hartree approximation. The linear entropy is defined as 
\begin{equation}\label{def:Impurity}
    S_L : = 1- \Tr\,\hat{\rho}^{(1)2}\,,
\end{equation}
and quantifies the closedness of $\rho^{(1)}$ to a pure state.
If $S_L\approx 0$, the one-particle marginal state is approximately pure, which implies that the $k$-particle marginal states can be approximated by a product state for any finite $k$: $\hat{\rho}^{(k)}\approx\hat{\rho}^{(1)\otimes k}$ for large $N$. Then, we define the infidelity with respect to the solution of the Hartree equation as 
\begin{equation}\label{def:Infidelity}
    I : = 1-\langle\phi|\hat{\rho}^{(1)}|\phi\rangle\,,
\end{equation}
which measures if such solution correctly describes the one-particle marginal state of the solution of the corresponding many-body dynamics.

The case where $A^{(2)}=0$ and the initial condition is a product state is a trivial example where  $S_L=0$ and $I=0$ at any time. For Hermitian Hamiltonians, it is known \cite{knowles2010mean} that the infidelity with respect to the solution of the Hartree equation scales as $N^{-1}$ for a broad family of two-body Hamiltonians.  However, we can already observe from equation \eqref{eq:partial_t_rho_exact_normalized} that the last two terms depending on the anti-Hermitian part of the Hamiltonian have a prefactor $\mathcal{O}(N)$. Then, even in the case where $\hat{\rho}^{(2)} = \hat{\rho}^{(1)\otimes 2} + \mathcal{O}\left(N^{-1}\right)$, the non-Hermitian terms in the Hamiltonian will generate corrections to the Hartree equation that remain finite in the limit $N\to\infty$. So, in the limit $N\to \infty$, the evolution of $\hat{\rho}^{(1)}$ will depend on $\hat{\rho}^{(2)}$ and $\hat{\rho}^{(3)}$, and this is not captured by the Hartree equation. In the next subsection we will show how this amplification due to the anti-Hermitian terms works at short times.

\subsection{Short-time behavior of the reduced one-particle state} \label{sec:derivatives}

In this subsection, we compare the first and second time derivatives at $t=0$ of the solution of the Hartree equation \eqref{eq:HF_non_Hermitian_normalized} and of the exact one-particle reduced density matrix \eqref{eq:partial_t_rho_exact_normalized}. We will show that, even when the initial state is a product state, the Hartree approximation captures the correct first time derivative but it fails to describe the second time derivative. This deviation is only present for Hamiltonians that are not Hermitian and contain interaction terms. 

The first derivatives of $\gamma$ and of $\hat{\rho}^{(1)}$ are given by the Hartree equation \eqref{eq:HF_non_Hermitian_normalized} and by the hierarchy \eqref{eq:partial_t_rho_exact_normalized}, respectively.  Observe that at $t=0$, since $\hat{\rho}^{(1)}(0)=\gamma(0)=|\varphi\rangle \langle\varphi|$, and furthermore since $\hat{\rho}^{(k)}(0)=\gamma(0)^{\otimes k}$, both expressions match:
\begin{equation}\label{eq:comparison_firstderivative}
    i \partial_t \gamma |_{t=0}=i \partial_t \hat{\rho}^{(1)} |_{t=0}.
\end{equation}
This is true by construction, because the last two terms in \eqref{eq:partial_t_rho_exact_normalized}, scaling with $N$, cancel exactly. 

Let us now consider the first time derivative of the $k$-particle reduced density matrices. By taking the partial trace of the evolution equation \eqref{eq:Many-body_Schrodinger} over the last $N-k$ particles, we can compute the evolution equation of the $k$-particle reduced density matrix. One can then normalize it and, proceeding as in the derivation of \eqref{eq:partial_t_rho_exact_normalized}, we get the following evolution equation:
\begin{equation}\label{eq:idt_rhok}
\begin{split}
    i \partial_t \hat{\rho}^{(k)} =& \sum_{i=1}^{k}\left( A^{(1)}_i - \Tr\left(A^{(1)}\hat{\rho}^{(1)}\right)\right)\hat{\rho}^{(k)} - h.c. \\
    &+ \frac{N-k}{N-1} \sum_{i=1}^{k} \Tr_{k+1}\left( \left( A^{(2)}-\Tr\left(A^{(2)} \hat{\rho}^{(2)}\right)\right)_{i,k+1} \hat{\rho}^{(k+1)}\right) -h.c. \\
    &+ (N-k) \Tr_{k+1} \left(G^{(1)}_{k+1}\left( \hat{\rho}^{(k+1)}-\hat{\rho}^{(k)}\otimes \hat{\rho}^{(1)}\right) \right) \\
    &+ \frac{1}{N-1}\binom{N-k}{2} \Tr_{k+1,k+2} \left( G^{(2)}_{k+1,k+2} \left( \hat{\rho}^{(k+2)}-\hat{\rho}^{(k)}\otimes \hat{\rho}^{(2)}\right) \right) \\
    &+\frac{1}{N-1} \sum_{1\leq i<j\leq k} \left(A^{(2)}-\Tr\left(A^{(2)}\hat{\rho}^{(2)}\right)\right)_{ij} \hat{\rho}^{(k)} - h.c.\,.
\end{split}
\end{equation}
In order to simplify notation, we define $G^{(i)}=A^{(i)}-A^{(i)\dagger}$ for $i=1,2$, we use sub indexes to indicate the particle where an operator acts and we write \emph{h.c.} for Hermitian conjugate. At time $t=0$, using that $\rho^{(k)}(0)=\gamma(0)^{\otimes k}$, the time derivative simplifies to

\begin{equation}\label{eq:idt_rhok-t0}
\begin{split}
    i \partial_t \hat{\rho}^{(k)}|_{t=0} =& \sum_{i=1}^{k}\left( A^{(1)}_i - \Tr\left(A^{(1)}\gamma(0)\right)\right)\gamma(0)^{\otimes k} -h.c. \\
    &+ \frac{N-k}{N-1} \sum_{i=1}^{k} \Tr_{k+1} \left(\left( A^{(2)}+\Tr\left(A^{(2)} \gamma(0)^{\otimes 2}\right)\right)_{i,k+1}  \gamma(0)^{\otimes k+1} \right)-h.c. \\
    &+\frac{1}{N-1} \sum_{1\leq i<j\leq k} \left(A^{(2)}-\Tr\left(A^{(2)}\gamma(0)^{\otimes 2}\right)\right)_{ij} \gamma(0)^{\otimes k} - h.c. \,.
\end{split}
\end{equation}
On the other hand, using the Hartree equation, one gets the time derivative
\begin{equation}\label{eq:idt_gammak}
\begin{split}
    i \partial_t \gamma^{\otimes k} = \sum_{i=1}^{k}\left( A^{(1)} - \Tr\left(A^{(1)}\gamma\right)\right)_i\ \gamma^{\otimes k} + \sum_{i=1}^{k} \Tr_{k+1}\left( \left( A^{(2)}+\Tr\left(A^{(2)} \gamma^{\otimes 2}\right)\right)_{i,k+1} \gamma^{\otimes k+1} \right) -h.c.
\end{split}
\end{equation}
Comparing \eqref{eq:idt_rhok-t0} and \eqref{eq:idt_gammak}, one can see that 
\begin{equation}\label{eq:comparison_k-firstderivative}
    i \partial_t \hat{\rho}^{(k)}|_{t=0} = i \partial_t \gamma^{\otimes k}|_{t=0} + \mathcal{O}\left(\frac{k}{N}\right).
\end{equation}
Then, for any fixed $k$, in the limit of large number of particles the Hartree equation describes correctly the first time derivative of $\hat{\rho}^{(k)}$. However, it is important to observe that the approximation error is $\mathcal{O}(N^{-1})$, because the next step is to compute the second derivative, and we will show that these small corrections are amplified in the non-Hermitian case and lead to a finite difference between the second derivative of the exact one-particle reduced density matrix and of the Hartree solution. Taking the time derivative of \eqref{eq:HF_non_Hermitian_normalized} one gets the second time derivative of the Hartree solution:
\begin{equation}\label{eq:idt2_gamma}
\begin{split}
    - \partial^2_t \gamma =& \left( A^{(1)} - \Tr\left( A^{(1)} \gamma \right) \right) i\partial_t\gamma  - \Tr\left(A^{(1)}i\partial_t\gamma\right)\gamma - h.c.\\
    &+\Tr_{2}\left( \left( A^{(2)}-\Tr\left(A^{(2)} \gamma^{\otimes 2}\right)\right) i\partial_t\gamma^{\otimes 2}\right) -  \Tr\left( A^{(2)} i\partial_t \gamma^{\otimes2} \right) \gamma  -h.c.\,.
\end{split}
\end{equation}

We compute the second time derivative of the one-particle reduced density matrix by taking the time derivative of \eqref{eq:partial_t_rho_exact_normalized}:
\begin{equation}\label{eq:partial^2_t_rho_exact_normalized}
\begin{split}
    - \partial_t^2 \hat{\rho}^{(1)} =& \left( A^{(1)} - \Tr\left(A^{(1)}\hat{\rho}^{(1)}\right) \right) i\partial_t\hat{\rho}^{(1)} - \Tr\left(A^{(1)}i\partial_t\hat{\rho}^{(1)}\right) \hat{\rho}^{(1)}  + h.c.\\
    &+ \Tr_{2}\left(\left(A^{(2)} - \Tr\left(A^{(2)}\hat{\rho}^{(2)}\right) \right) i\partial_t\hat{\rho}^{(2)} \right) - \Tr\left(A^{(2)}i\partial_t\hat{\rho}^{(2)}\right) \hat{\rho}^{(1)} +h.c. \\
    &+\left(N-1\right) \Tr_{2}\left(G^{(1)}_2 \left(i\partial_t \hat{\rho}^{(2)} - i\partial_t \hat{\rho}^{(1)} \otimes \hat{\rho}^{(1)}- \hat{\rho}^{(1)}\otimes i\partial_t\hat{\rho}^{(1)} \right)\right) \\
    &+ \frac{N-2}{2} \Tr_{23}\left(G^{(2)}_{23} \left(i\partial_t\hat{\rho}^{(3)} - i\partial_t\hat{\rho}^{(1)} \otimes \hat{\rho}^{(2)} - \hat{\rho}^{(1)} \otimes i\partial_t\hat{\rho}^{(2)}\right)  \right).
\end{split}
\end{equation}

Using the previous results \eqref{eq:comparison_firstderivative} and \eqref{eq:comparison_k-firstderivative}, it follows that the first two lines in \eqref{eq:partial^2_t_rho_exact_normalized} are reproduced by \eqref{eq:idt2_gamma} at the initial time up to small corrections $\mathcal{O}(N^{-1})$. Now we will compute the last two lines and show that they contribute with extra terms which modify the second time derivative. 

Let us use \eqref{eq:idt_rhok-t0} with $k=1,2$, to compute 
\begin{equation}\label{eq:rho2diff}
\begin{split}
    &\left(i\partial_t \hat{\rho}^{(2)} - i\partial_t \hat{\rho}^{(1)} \otimes \hat{\rho}^{(1)}+ \hat{\rho}^{(1)}\otimes i\partial_t\hat{\rho}^{(1)}\right)\Big|_{t=0} \\
    &= -\frac{1}{N-1}\sum_{i=1}^{2} \Tr_3\left( \left(A^{(2)}-\Tr\left(A^{(2)}\gamma(0)^{\otimes2}\right)\right)_{i,3} \gamma(0)^{\otimes3}\right) -h.c. \\
    &\hspace{1em}+ \frac{1}{N-1} \left(A^{(2)}- \Tr\left(A^{(2)}\gamma(0)^{\otimes 2}\right) \right)\gamma(0)^{\otimes 2} - h.c.\,.
\end{split}
\end{equation}
Similarly, we can use \eqref{eq:idt_rhok-t0} with $k=1,2,3$ to compute
\begin{equation}\label{eq:rho3dif}
\begin{split}
    &\left(i\partial_t\hat{\rho}^{(3)} - i\partial_t\hat{\rho}^{(1)} \otimes \hat{\rho}^{(2)} + \hat{\rho}^{(1)} \otimes i\partial_t\hat{\rho}^{(2)}\right) \Big|_{t=0} \\
    &= -\frac{1}{N-1} \sum_{i=2}^{3} \Tr_4\left( \left(A^{(2)}-\Tr\left(A^{(2)} \gamma(0)^{\otimes2}\right)\right)_{i4}\gamma(0)^{\otimes4}\right) -h.c.\\
    &\hspace{1em} - \frac{2}{N-1} \Tr_4\left( \left(A^{(2)}-\Tr\left(A^{(2)} \gamma(0)^{\otimes2}\right)\right)_{14}\gamma(0)^{\otimes4}\right) -h.c.\\
    &\hspace{1em} +\frac{1}{N-1}\sum_{i=2}^{3} \left(A^{(2)}-\Tr\left(A^{(2)} \gamma(0)^{\otimes2}\right)\right)_{1i}\gamma(0)^{\otimes3}-h.c.\,.
\end{split}
\end{equation}
Plugging expressions \eqref{eq:rho2diff} and \eqref{eq:rho3dif} into \eqref{eq:partial^2_t_rho_exact_normalized}, subtracting the second derivative of the Hartree equation at time $t=0$  and using the properties of the partial trace, we get

\begin{equation}\label{eq:diff_secondderivatives}
\begin{split}
    -\partial^2_t(\hat{\rho}^{(1)}-\gamma)|_{t=0} =& \Bigg[\Tr_2\left( \left( G^{(1)} - \Tr\left( G^{(1)} \gamma(0) \right) \right)_2 A^{(2)} \gamma(0)_2 \right) \\
    &\hspace{1em}- \Tr\left( G^{(1)}_2 \left(A^{(2)}-\Tr\left(A^{(2)}\gamma(0)^{\otimes 2}\right)\right) \gamma(0)^{\otimes 2} \right)\Bigg] \gamma(0) + h.c.\\
    &+\Bigg[\Tr_{23}\left( \left(G^{(2)}-\Tr\left(G^{(2)}\gamma(0)^{\otimes 2}\right)\right)_{23} A^{(2)}_{12} \gamma(0)_{2}\gamma(0)_{3}\right) \\
    &\hspace{1em}-\Tr\left(G^{(2)}_{23}\left(A^{(2)}-\Tr\left(A^{(2)}\gamma(0)^{\otimes 2}\right)\right)_{12}\gamma^{\otimes3}\right) \Bigg] \gamma(0) + h.c. \\
    &+\mathcal{O}\left(\frac{1}{N}\right)
\end{split}
\end{equation}

The conclusion of this computation is that the Hartree equation is constructed in such a way that at $t=0$, when the system is exactly in a product state, the first time derivative of the Hartree solution and of the exact one-particle reduced density matrix are the same, and this is shown in \eqref{eq:comparison_firstderivative}. The same construction shows that for any fixed $k$, the difference between the time derivative of $k$ copies of the Hartree solution and the $k$-particle reduced density matrix vanishes for large $N$ with an error of order $\mathcal{O}\left(N^{-1}\right)$ as shown in \eqref{eq:comparison_k-firstderivative}. Even if the first time derivative matches, the difference between the second derivative of the one-particle reduced density matrix and of the Hartree solution has contributions that do not vanish for large $N$, and this is shown in \eqref{eq:diff_secondderivatives}. This contributions are present only on the non-Hermitian case, since they depend on the anti-hermitian Hamiltonian terms $G^{(i)}$, $i=1,2$ and furthermore they also vanish if there is no interaction, i.e., if $A^{(2)}=0$.  This shows how the non-Hermitian terms of the Hamiltonian amplify the approximation error of the mean-field approximation in such a way that even at small times, the difference between the Hartree solution and the exact one-particle reduced density matrix does not vanish.

\section{ First counterexample: purely anti-Hermitian Hamiltonian}
\label{sec:ZZ_hamiltonian}

We present a nontrivial example in which the solution of the non-Hermitian nonlinear Schr\"odinger equation \eqref{eq:HF_non_Hermitian_normalized} and the limit for $N\to\infty$ of the one-particle normalized marginal state \eqref{eq:Many-body_Schrodinger_Psi} can both be analytically computed and do not agree.

Let each particle be a qubit with Hilbert space $\h=\mathbb{C}^{2} = \mathrm{Span}\{\ket{0},\ket{1}\}$. We write the initial one-particle wavefunction $\varphi\in \h$ in the computational basis: $\ket{\varphi} = \varphi_0\ket{0} + \varphi_1 \ket{1}$.
Let us choose $A^{(1)}=0$ and $A^{(2)}=iZ\otimes Z$, where $Z = \ket{0}\bra{0}- \ket{1}\bra{1}$ is the third Pauli matrix. Similar anti-Hermitian interactions can be engineered in qubit systems \cite{zeni2025}. 
We choose the following orthonormal basis of $\text{Sym}^N\mathbb{C}^{2}$, indexed by $n=0,...,N$:
\begin{equation}
    |N,n\rangle := \frac{1}{N!}\sqrt{\binom{N}{n}}\sum_{P \in S_N} P\left(\ket{0}^{\otimes n} \otimes \ket{1}^{\otimes (N-n)}\right)\,,
\end{equation}
where the sum runs over all the $N!$ permutations of $N$ qubits.
This basis diagonalizes the evolution operator:
\begin{equation}\label{eq:eigenvector}
    -iA^{(N)} |N,n\rangle  = \lambda_{N,n} |N,n\rangle,\quad \lambda_{N,n} = \frac{(N-2n)^2 - N}{2(N-1)}\,.
\end{equation}
In this basis, the initial $N$-particle wavefunction has the form
\begin{equation}
    |\Psi^{(N)}(0)\rangle = \left( \varphi_0 \ket{0} + \varphi_1 \ket{1} \right)^{\otimes N} = \sum_{n=0}^N  \varphi_0^{n}\varphi_1^{N-n}\sqrt{\binom{N}{n}} |N,n\rangle\,,
\end{equation}
and the solution of the evolution equation \eqref{eq:Many-body_Schrodinger} becomes 
\begin{equation}\label{eq:Psi_N(t)}
    |\Psi^{(N)}\rangle =  \sum_{n=0}^N  \varphi_0^{n}\varphi_1^{N-n}\sqrt{\binom{N}{n}} e^{\lambda_{N,n}t} |N,n\rangle.
\end{equation}
To compute the unnormalized one-particle marginal state, we can compute the partial trace using the basis $\{\ket{N-1,s}\}_{s=0}^{N-1}$ of $\mathrm{Sym}^{N-1}\mathbb{C}^2$:
\begin{equation}\label{eq:rho^{(1)}_def}
    \rho^{(1)} = \sum_{s=0}^{N-1} \langle N-1,s|\Psi^{(N)}\rangle\langle\Psi^{(N)}|N-1,s\rangle\,.
\end{equation}
We have
\begin{equation}\label{eq:sigmA^{(N)}-sigmA^{(N)}-1}
\begin{split}
    &\langle N-1,s|N,n\rangle \\
    &=\sum_{P\in S^N}\sum_{Q\in S^{N-1}} \frac{1}{N!(N-1)!}\sqrt{\binom{N}{n}\binom{N-1}{s}} Q\left(\bra{0}^{\otimes s} \otimes \bra{1}^{\otimes(N-1-s)} \right) P\left(\ket{0}^{\otimes n} \otimes \ket{1}^{\otimes(N-n)} \right) \\
    &= \sum_{P\in S^N} \frac{1}{N!}\sqrt{\binom{N}{n}\binom{N-1}{s}} \left(\bra{0}^{\otimes s} \otimes \bra{1}^{\otimes(N-1-s)} \right) P\left(\ket{0}^{\otimes n} \otimes \ket{1}^{\otimes(N-n)} \right) \\
    &= \frac{1}{N!}\sqrt{\binom{N}{n}\binom{N-1}{s}}n!(N-n)!  \Big( \delta_{n,s} \ket{1} + \delta_{n,s+1} \ket{0} \Big) \\
    & = \sqrt{\frac{\binom{N-1}{s}}{\binom{N}{n}}}  \left( \delta_{n,s} \ket{1} + \delta_{n,s+1} \ket{0} \right)\,.
\end{split}
\end{equation}

Plugging equation \eqref{eq:Psi_N(t)} in \eqref{eq:rho^{(1)}_def} and using \eqref{eq:sigmA^{(N)}-sigmA^{(N)}-1} we obtain the matrix elements of $\rho^{(1)}$:
\begin{equation}\label{eq:rho_entries}
\begin{split}
    \rho^{(1)}_{00} &= |\varphi_0|^2 \sum_{s=0}^{N-1} |\varphi_1|^{2 s} |\varphi_0|^{2(N-1-s)} \binom{N-1}{s} e^{2\lambda_{N,s}t}\,, \\
    \rho^{(1)}_{01} &= \varphi_0 \varphi_1^* \sum_{s=0}^{N-1} |\varphi_0|^{2 s} |\varphi_1|^{2(N-1-s)} \binom{N-1}{s} e^{(\lambda_{N,s+1}+\lambda_{N,s})t}\,, \\
    \rho^{(1)}_{10} &= \varphi_1\varphi_0^* \sum_{s=0}^{N-1} |\varphi_0|^{2 s} |\varphi_1|^{2(N-1-s)} \binom{N-1}{s} e^{(\lambda_{N,s+1}+\lambda_{N,s})t}\,, \\
    \rho^{(1)}_{11} &= |\varphi_1|^2 \sum_{s=0}^{N-1} |\varphi_0|^{2 s} |\varphi_1|^{2(N-1-s)} \binom{N-1}{s} e^{2\lambda_{N,s}t}\,. \\
\end{split}
\end{equation}
To analyze the evolution equation of the Hartree wavefunction \eqref{eq:HF_non_Hermitian_vector}, we write the wavefunction in the computational basis $\ket{\phi} = \phi_0 \ket{0}+ \phi_1 \ket{1}$ with initial condition $\ket{\phi(0)}=\ket{\varphi}$ and get the equations
\begin{equation} \label{eq:HF_ZZ}
    \begin{cases}
        \partial_t \ket{\phi} = \left(\bra{\phi}Z\ket{\phi} Z - \bra{\phi}Z\ket{\phi}^2 \right) |\phi\rangle \\
        \ket{\phi(0)} = \ket{\varphi} 
    \end{cases}
    \Rightarrow
    \begin{cases}
    \partial_t \phi_0 = \left(  |\phi_0|^2 - |\phi_1|^2 - \left(  |\phi_0|^2 - |\phi_1|^2 \right)^2\right) \phi_0\\
    \partial_t \phi_1 = \left(  |\phi_1|^2 - |\phi_0|^2  - \left(  |\phi_0|^2 - |\phi_1|^2 \right)^2\right) \phi_1 \\
    \quad\phi_0(0) = \varphi_0,\quad \phi_1(0) = \varphi_1
    \end{cases}
\end{equation}
The solution to these equations is
\begin{equation}\label{eq:HF_solution}
    \phi_0 =e^{i\theta_0} \sqrt{\frac{1}{2}\left( 1+ \frac{k e^{2t}}{\sqrt{1+(ke^{2t})^2}}\right)}, \quad \phi_1 = e^{i\theta_1}\sqrt{\frac{1}{2}\left( 1- \frac{k e^{2t}}{\sqrt{1+(ke^{2t})^2}}\right)}\,,
\end{equation}
where
\begin{equation}\label{def:k,theta}
 k := \frac{|\varphi_0|^2-|\varphi_1|^2}{\sqrt{1-(|\varphi_0|^2-|\varphi_1|^2)^2}},\quad \theta_{i} = \text{arg}(\varphi_i)\,.
\end{equation}

Equation \eqref{eq:HF_ZZ} has stable and unstable fixed points. Wavefunctions with $|\varphi_0|^2=|\varphi_1|^2=\frac{1}{2}$ are unstable fixed points, while the stable fixed points are given by $\ket{\varphi}=\ket{0}$ and $\ket{\varphi}=\ket{1}$.

We will show that the system can have the following three asymptotic behaviors for $N\to\infty$:
\begin{itemize}
    \item Stable fixed points: if $|\varphi\rangle=|0\rangle$ or $|\varphi\rangle=|1\rangle$, the solution of the Hartree equation is constant in time and agrees with the one-particle marginal state for any $N$.
    \item Unstable fixed points after the critical time: if $|\varphi_0|^2 = |\varphi_1|^2 = \frac{1}{2}$, the solution of the Hartree equation is constant in time and agrees up to $t=\frac{1}{2}$ with the limit for $N\to\infty$ of the one-particle marginal state.
    For $t>\frac{1}{2}$, such a limit becomes mixed, while the solution of the Hartree equation stays constant.
    \item Any other case: if $|\varphi_0| \neq |\varphi_1|$, the limit for $N\to\infty$ of the one-particle marginal state is pure but is different from the solution of the Hartree equation.
\end{itemize}

\subsection{Large-\texorpdfstring{$N$}{N} limit}
\label{sec:Large_N_limit}
In this section, we will compute the limit of the one-particle marginal state $\hat{\rho}^{(1)}$ for $N\to\infty$.

Let us first consider the initial conditions that are stable fixed points of the Hartree equation.
We will consider only the case $|\varphi\rangle = |0\rangle$, since the case $|\varphi\rangle = |1\rangle$ is completely analogous.
The solution to the $N$-body evolution equation \eqref{eq:Psi_N(t)} is $\ket{\Psi_N} = e^{\frac{Nt}{2}}\ket{0}^{\otimes N}$ and the one-particle marginal state is constant in time: $\hat{\rho}^{(1)}=\ket{0}\bra{0}$. The linear entropy and the infidelity with respect to the Hartree solution are $S_L=0$, $I=0$ and the Hartree approximation is exact at any time for any $N$.

In the remainder of this section, we will consider the initial conditions satisfying $|\varphi_0|^2,\,|\varphi_1|^2\in(0,1)$.
We will use the Stirling's approximation of the factorial
\begin{equation}
    \ln M! = M\ln M - M +\frac{1}{2}\ln\left(2\pi M\right)+ \mathcal{O}(M^{-1}).
\end{equation}
Then we can approximate the binomial distribution as
\begin{equation}\label{eq:binomial_stirling}
    p^n q^{M-n}\binom{M}{n} \approx \exp\left(n\ln\frac{p\left(M-n\right)}{q\,n} +M\ln\frac{q\,M}{M-n} + \frac{1}{2}\ln\frac{M}{2\pi n(M-n)} \right)\,.
\end{equation}
We will use the substitution $s=Nx$ such that we can approximate the sum over $s$ by an integral over $x$. We can also approximate the eigenvalues as
\begin{equation}\label{eq:approx_lambda}
    \lambda_{N,Nx} = \frac{N}{2} (1-2x)^2 -2x(1-x) + \mathcal{O}(N^{-1}).
\end{equation}
We rewrite $\rho^{(1)}_{00}$ as 
\begin{equation}
    \rho^{(1)}_{00} = \sum_{s=0}^{N}|\varphi_1|^{2s} |\varphi_0|^{2(N-s)} \frac{N-s}{N}\binom{N}{s} e^{2\lambda_{N,s}t}.
\end{equation}
Applying the Stirling's approximation \eqref{eq:binomial_stirling}, replacing $s=Nx$, using \eqref{eq:approx_lambda} and approximating the sum by an integral, we get the following expression: 
\begin{equation}\label{eq:rho_int}
    \rho_{00}^{(1)} \approx \sqrt{\frac{N}{2\pi}}\int_0^1 dx\ \sqrt{\frac{1-x}{x}} e^{N f_t(x)-4tx(1-x)}\,,
\end{equation}
where
\begin{equation}\label{eq:f}
    f_t(x) = x \ln\frac{|\varphi_1|^2\left(1-x\right)}{|\varphi_0|^2\,x} +\ln\frac{|\varphi_0|^2}{1-x} + t(1-2x)^2\,,
\end{equation}
and its derivatives are
\begin{equation}\label{eq:f'}
    f'_t(x) = \ln\frac{|\varphi_1|^2}{|\varphi_0|^2} + \ln\frac{1-x}{x} - 4t(1-2x)\,,
\end{equation}
\begin{equation}\label{eq:f''}
    f''_t(x) = 8t - \frac{1}{x(1-x)}\,.
\end{equation}

The integral \eqref{eq:rho_int} can be further approximated using the Laplace's method \cite{murray1984asymptotic, copson2004asymptotic}. In the limit of large $N$, the major contributions to the integral come from the values of $x$ near the global maximizers of $f_t$. In the neighborhood of these points, we can approximate the integrand by a Gaussian function using a Taylor expansion for $f_t$. Since the tails of these integrands decay exponentially, we can extend the integral to the whole real line. In this way we can integrate analytically the Gaussian function. Let $\{x_i\}$ be the global maximizers of $f_t$. Recall that $|\varphi_0|^2,|\varphi_1|^2\in(0,1)$. The derivative \eqref{eq:f'} satisfies $\lim_{x\to 0^{+}}f'_t(x)=+\infty$ and $\lim_{x\to 1^{-}}f'_t(x)=-\infty$. Then, all the global maximizers of $f_t$ lie in $(0,1)$ and satisfy  $f'_t(x_i)=0$, $f''_t(x_i)\le0$. The Laplace's method applied to equation \eqref{eq:rho_int} gives
\begin{equation}
    \rho_{00}^{(1)} \approx \sum_i \sqrt{\frac{N}{2\pi}} \sqrt{\frac{1-x_i}{x_i}} e^{Nf_t(x_i)-4tx_i(1-x_i)} \int_{-\infty}^{\infty} dx\ e^{\frac{N}{2}f''(x_i)(x-x_i)^2}.
\end{equation}
Using the Gaussian integral  $\int_{-\infty}^{\infty} e^{-y^2} dy = \sqrt{\pi}$, the result is  
\begin{equation}\label{eq:rho_laplace}
    \rho_{00}^{(1)} \approx \sum_i \sqrt{\frac{1}{-f''(x_i)}} \sqrt{\frac{1-x_i}{x_i}} e^{Nf_t(x_i)-4tx_i(1-x_i)}\ .
\end{equation}

\begin{figure}[h!t]
\centering
\begin{subfigure}{0.49\textwidth}
    \includegraphics[width=\textwidth]{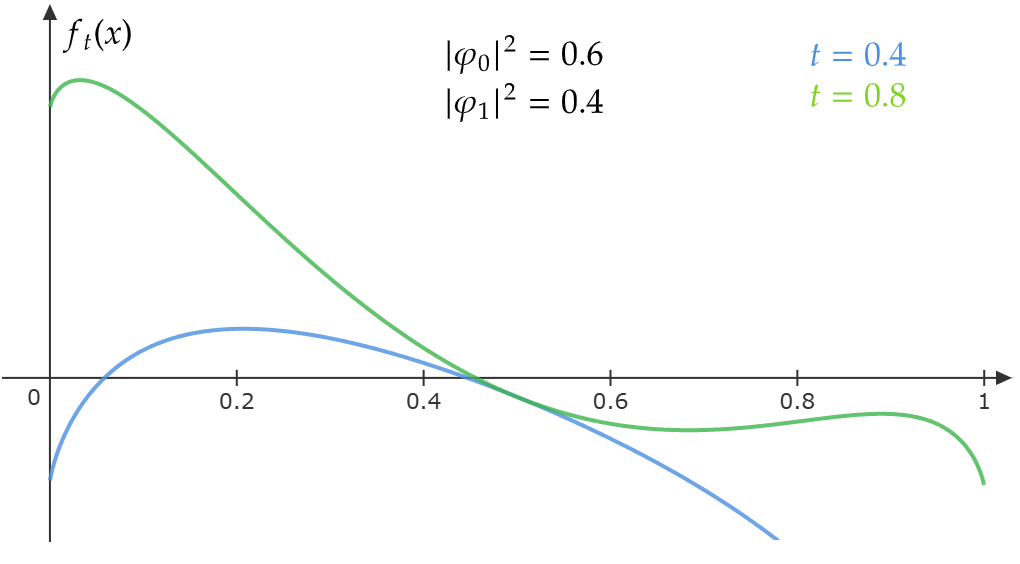}
    \caption{For initial conditions satisfying $|\varphi_0|^2\neq|\varphi_1|^2$, the function $f_t$ can have one or two local maximizers, but it always has a single global maximizer.}
    \label{fig:f(x)04}
\end{subfigure}
\hfill
\begin{subfigure}{0.49\textwidth}
    \includegraphics[width=\textwidth]{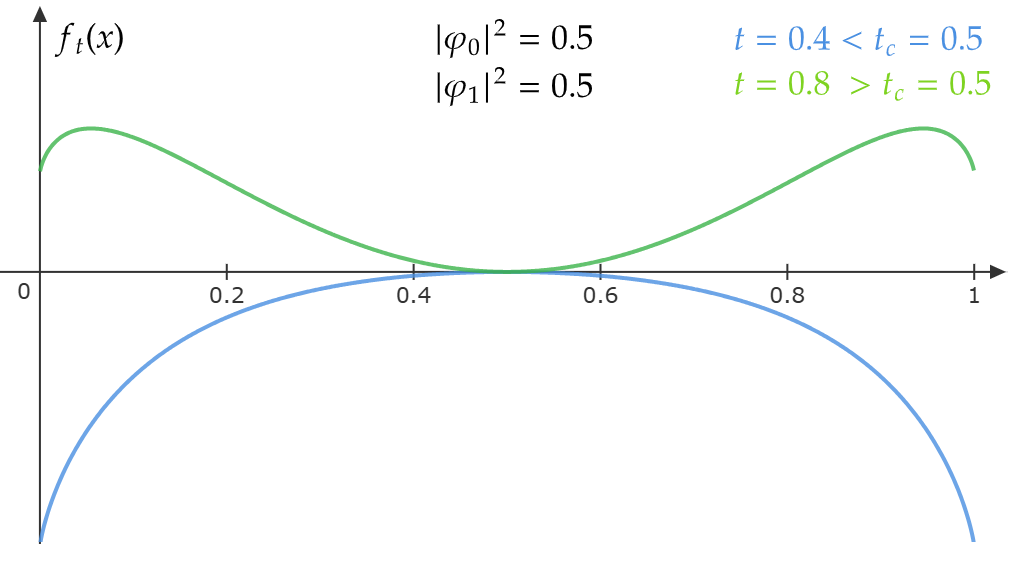}
    \caption{For the initial conditions satisfying $|\varphi_0|^2=|\varphi_1|^2=\frac{1}{2}$, the function $f_t$ has a single local maximizer $x=\frac{1}{2}$ for $t\le\frac{1}{2}$ and it has two symmetric global maximizers for $t>\frac{1}{2}$.}
    \label{fig:f(x)05}
\end{subfigure}
\caption{{\bf Purely anti-Hermitian Hamiltonian:} Plot of $f_t$ as a function of $x$, as defined in \eqref{eq:f}, for different instances of initial condition and time.}
\label{fig:f(x)}
\end{figure}

We will show that for any initial condition with $|\varphi_0|^2\neq|\varphi_1|^2$, $f_t$ has a single global maximizer. In \autoref{fig:f(x)04} we plot the function $f_t$ for a generic initial condition. $f_t$ may have one or two local maximizers, but only one of them is a global maximizer.  Only for the exceptional initial condition $|\varphi_0|^2=|\varphi_1|^2=\frac{1}{2}$ and $t>\frac{1}{2}$ there are two global maximizers (see \autoref{fig:f(x)05}).

To prove the uniqueness of the global maximizer, we start by analyzing the function $f''_t(x)$ given by equation \eqref{eq:f''}.
At times $t<\frac{1}{2}$, the second derivative is always negative: $f''_t(x)<0\ \forall\;x\in(0,1)$. Hence $f'_t$ is strictly decreasing and $f_t$ has a single local maximizer, which is then the unique global maximizer. 
At times $t>\frac{1}{2}$, since equation \eqref{eq:f''} is symmetric with respect to $x=\frac{1}{2}$, $f''_t$ has two symmetric roots
\begin{equation}\label{eq:roots}
    r_{\pm} = \frac{1}{2}\left(1\pm\sqrt{1-\frac{1}{2t}}\right),
\end{equation}
which split the interval $(0,1)$ into three sub-intervals. $f''_t$ is positive in the central sub-interval and negative in the other two. When $f_t$ has two local maximizers, one of them is larger than $r_+$ and the other one is smaller than $r_-$. To determine which one is a global maximizer, we observe that if $x^*_t$ is a global maximizer then
\begin{equation}
    f_t(x_t^*)\geq f_t(1-x^*_t)= f_t(x^*_t) + (1-2x^*_t)\ln\frac{|\varphi_1|^2}{|\varphi_0|^2}\,,
\end{equation}
and
\begin{equation}\label{eq:max_cond}
    (1-2x^*_t)\ln\frac{|\varphi_1|^2}{|\varphi_0|^2} \leq 0\,.
\end{equation}
If $|\varphi_0|^2>|\varphi_1|^2$, the unique global maximizer satisfies $x_t^*< r_-$. If $|\varphi_0|^2<|\varphi_1|^2$, the unique global maximizer satisfies $x_t^*>r_+$.
If $|\varphi_0|^2=|\varphi_1|^2=\frac{1}{2}$, the equation \eqref{eq:max_cond} is trivially satisfied and there are two global maximizers wich are symmetric with respect to $\frac{1}{2}$.

One can compute the evolution equation of the global maximizer implicitly using $f'_t(x^*_t)=0$:
\begin{equation}\label{eq:dt_x^*}
    \partial_tx^*_t = \frac{4(1-2x^*_t)}{8t - \frac{1}{x^*_t(1-x^*_t)}}.
\end{equation}
At time $t<\frac{1}{2}$ the denominator is strictly negative. At time $t>\frac{1}{2}$ the denominator vanishes if the equation \eqref{eq:f''} vanishes. If a point satisfies  $f'_t(x)=0$ and $f''_t(x)=0$, then it is a saddle point and it is not the global maximizer, unless $f'''_t(x)=0$, which only happens at $x=\frac{1}{2}\in(r_-,r_+)$. Then, the denominator of \eqref{eq:dt_x^*} never vanishes, and the trajectory of the global maximizer is continuous. 

At the critical time $t_c=\frac{1}{2}$, the trajectory can be discontinuous only if the maximizer is $x_{\frac{1}{2}}^*=\frac{1}{2}$ and this happens only in the exceptional case $|\varphi_0|^2=|\varphi_1|^2$. In this case, $x_t^*=\frac{1}{2}$ at $t\leq\frac{1}{2}$ and at later times there are two global maximizers.

\subsubsection{Single global maximizer}
\label{sec:single_maxima}
For the initial conditions such that $|\varphi_0|^2\neq|\varphi_1|^2$ or $|\varphi_0|^2=|\varphi_1|^2$ and $t\leq \frac{1}{2}$, we showed that there is a unique global maximizer $x^*_t$. Equation \eqref{eq:rho_laplace} becomes
\begin{equation}\label{eq:lim_N_rho(1)_00}
    \rho_{00}^{(1)} \approx \frac{1}{\sqrt{-f''_t(x^*_t)}} \sqrt{\frac{1-x^*_t}{x^*_t}} \exp\Bigg(N f_t(x^*_t)-4tx^*_t(1-x^*_t) \Bigg)\,.
\end{equation}
A similar computation allows us to compute the remaining matrix elements:
\begin{equation}
    \rho_{11}^{(1)} \approx \frac{1}{\sqrt{-f''_t(x^*_t)}} \sqrt{\frac{x^*_t}{1-x^*_t}} \exp\Bigg(N f_t(x^*_t)-4tx^*_t(1-x^*_t) \Bigg) \,,
\end{equation}
\begin{equation}
    \rho_{10}^{(1)} =\rho_{01}^{(1)\ *} \approx \frac{\varphi_0^*}{\varphi_1^*} \frac{1}{\sqrt{-f''_t(x^*_t)}} \sqrt{\frac{x^*_t}{1-x^*_t}}\exp\Bigg(N f_t(x^*_t)- 4tx^*_t(1-x^*_t) + 2t(1-2x^*_t)\Bigg)\,.
\end{equation}
When normalizing the state, all the $N$ dependence cancels and we get the limit of the one-particle marginal state for the case where $f_t$ has a single global maximizer:
\begin{equation} \label{eq:lim_N_rho(1)}
    \lim_{N\to \infty}\hat{\rho}^{(1)} = \begin{pmatrix}
        1-x^*_t & e^{i\theta} \sqrt{x^*_t(1-x^*_t)} \\
        e^{-i\theta} \sqrt{x^*_t(1-x^*_t)} & x^*_t 
    \end{pmatrix}\,,
\end{equation}
where $e^{-i\theta} = \frac{\varphi_0^*}{|\varphi_0|}\frac{|\varphi_1|}{\varphi_1^*} $. Observe that the limit state \eqref{eq:lim_N_rho(1)} is pure and can be characterized by a one-particle wavefunction: $\lim_{N\to \infty}\hat{\rho}^{(1)}=:\ket{\nu}\bra{\nu}$, where $\ket{\nu}= \nu_0 \ket{0} +  \nu_1 \ket{1}$ defined up to a global phase by
\begin{equation}\label{eq:nu_components}
    \nu_0= e^{i\theta}\sqrt{1-x^*_t},\quad \nu_1= \sqrt{x^*_t}\ .
\end{equation}
In order to compare it with the Hartree evolution, we derive the evolution equation for $\ket{\nu}$ using the implicit time evolution equation \eqref{eq:dt_x^*} for the global maximizer and we write it in terms of its components using equation \eqref{eq:nu_components}. The evolution equation for the one-particle wavefunction $\ket{\nu}$ is 

\begin{equation}\label{eq:lim_dtdnu}
\begin{cases}\displaystyle
    \partial_t \nu_0 = \frac{|\nu_0|^2-|\nu_1|^2 - \left(|\nu_0|^2-|\nu_1|^2\right)^2}{1-8t|\nu_0|^2|\nu_1|^2}\nu_0\\
    \displaystyle
    \partial_t \nu_1 = \frac{|\nu_1|^2-|\nu_0|^2 - \left(|\nu_0|^2-|\nu_1|^2\right)^2}{1-8t|\nu_0|^2|\nu_1|^2}\nu_1\,,
\end{cases}
\end{equation}
and matches the Hartree equation \eqref{eq:HF_ZZ} only at $t=0$ or at the fixed points.
This is a counterexample to the generalization of the mean-field theory to non-Hermitian Hamiltonians.

\subsubsection{Double global maximizer}
\label{sec:double_maxima}

For initial conditions such that $|\varphi_0|^2=|\varphi_1|^2=\frac{1}{2}$, for $t>\frac{1}{2}$ $f_t$ has two symmetric global maximizers $x_t^*\in(0,\frac{1}{2})$ and $1- x_t^*\in(\frac{1}{2},1)$. Evaluating the approximation \eqref{eq:rho_laplace} we obtain
\begin{equation}\label{eq:lim_N_rho(1)_00_exceptional}
    \rho_{00}^{(1)}  \approx \frac{1}{\sqrt{-f''_t(x^*_t)}} \Bigg(\sqrt{\frac{1-x^*_t}{x^*_t}} + \sqrt{\frac{x^*_t}{1-x^*_t}} \Bigg) \exp\Bigg(N f_t(x^*_t)-4tx^*_t(1-x^*_t) \Bigg).
\end{equation}
By symmetry we know that $\rho_{11}^{(1)}=\rho_{00}^{(1)}$.
Using the same methods we can compute the non diagonal terms
\begin{equation}
    \rho_{01}^{(1)} =\rho_{10}^{(1)*} \approx \frac{2 e^{i\theta}}{\sqrt{-f''_t(x^*_t)}} \exp\Bigg(N f_t(x^*_t)-4tx_t^*(1-x^*_t) \Bigg).
\end{equation}
Normalizing by the trace we get the limit of the one-particle marginal state in the case of two global maximizers:
\begin{equation}\label{eq:lim_N_rho(1)_exceptional}
\lim_{N\to \infty} \hat{\rho}^{(1)}= \begin{pmatrix}
        \frac{1}{2} & e^{i\theta} \sqrt{x^*_t(1-x^*_t)} \\
        e^{-i\theta} \sqrt{x^*_t(1-x^*_t)} &\frac{1}{2} \\
    \end{pmatrix}.
\end{equation}
The linear entropy of this matrix is
\begin{equation}\label{eq:S_L_inf}
    S_L\left(\displaystyle \lim_{N\to \infty}\hat{\rho}^{(1)}\right) = \frac{1}{2}- 2 x^*_t(1-x^*_t) > 0\,,
\end{equation}
hence the limiting state is mixed and cannot be described by the solution of the Hartree equation.
Furthermore, in the limit $t\to\infty$ we have $0<x_t^*<r_- \to0$, and the limiting one-particle state tends to the maximally mixed state for $t\to\infty$.
We also notice that the limit for $N\to\infty$ of the one-particle marginal state \eqref{eq:lim_N_rho(1)_exceptional} can be written as the mixture of two pure states that come from the contribution of the two maximizers of $f_t$.

\subsection{Analysis of the second time derivative}
In this subsection we use the equation \eqref{eq:diff_secondderivatives} for the second time derivative of the difference between the one-particle reduced density matrix and the Hartree solution and we will show that this equation can tell us the initial conditions for which the approximation works. 

In the present example, equation \eqref{eq:diff_secondderivatives} can be written in the form
\begin{equation}
    \lim_{N\to\infty}\partial_t^2(\hat{\rho}^{(1)}-\gamma)|_{t=0} = 2 \langle\varphi| Z|\varphi\rangle \left(1-\langle\varphi| Z|\varphi\rangle^2\right)\left(Z-\langle\varphi| Z|\varphi\rangle\right)|\varphi\rangle\langle\varphi| +h.c.\ .
\end{equation}
There are only two types of initial conditions for which this equation vanishes. One can see that the term $\langle\varphi| Z|\varphi\rangle$ vanishes for the exceptional points $|\varphi_0|=|\varphi_1|$  and the term $\left(1-\langle\varphi| Z|\varphi\rangle^2\right)$ vanishes for the eigenvectors of the $Z$ operator: $|\varphi_0|=0,|\varphi_1|=1$ and $|\varphi_0|=1,|\varphi_1|=0$. Also, the vector equation $\left(Z-\langle\varphi| Z|\varphi\rangle\right)|\varphi\rangle=0$ vanishes only in the eigenvector of the $Z$ operator. 

The condition $\partial_t^2(\hat{\rho}^{(1)}-\gamma)|_{t=0}=\mathcal{O}(N^{-1})$ is a necessary condition for the success of the Hartree approximation. However, in this example, it gives us the only initial conditions for which the approximation works. In the case of the eigenvalues of the $Z$ operators, the initial state is a stable fixed point of the Hartree equation and the many body state is an eigenvalue of the many body Hamiltonian, then the evolution is trivial in both cases and the approximation is exact. In the case of the exceptional points, the initial state is an instable fixed point of the Hartree equation and despite the non trivial evolution of the many body equation, the one-particle reduced density matrix is time independent up to the critical time, in this time interval the approximation works. 

\subsection{Numerical Analysis}
\label{sec:Numerical_Analysis}

In the previous section, we have analytically computed the one-particle marginal state in the limit of infinitely many particles. We have found that for almost every initial condition, such limit is pure but does not satisfy the non-Hermitian Hartree equation.
For the exceptional initial conditions $|\varphi_0|^2 = |\varphi_1|^2 = \frac{1}{2}$, there is a critical time after which the limiting state becomes mixed. In this section, we will compare the solution of the non-Hermitian Hartree equation \eqref{eq:HF_solution} with the one-particle marginal state for finite $N$ \eqref{eq:rho_entries}.

\subsubsection{Linear entropy of the one-particle marginal state}
\begin{figure}[h!t]
\centering
\begin{subfigure}[t]{0.49\textwidth}
    \includegraphics[height=5.5cm]{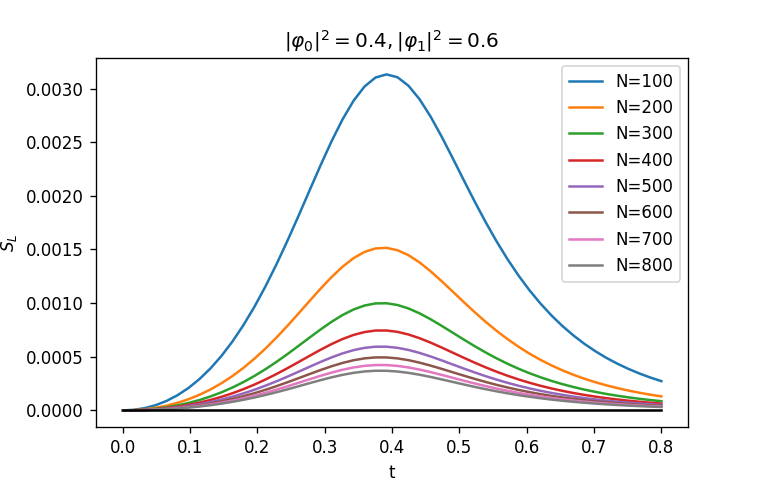}
    \caption{Initial conditions with $|\varphi_0|^2\neq|\varphi_1|^2$. Time is plotted in the horizontal axis. At any time, the linear entropy tends to zero for $N\to\infty$.}
    \label{fig:Impurity_t_at_04}
\end{subfigure}
\hfill
\begin{subfigure}[t]{0.49\textwidth}
    \includegraphics[height=5.5cm]{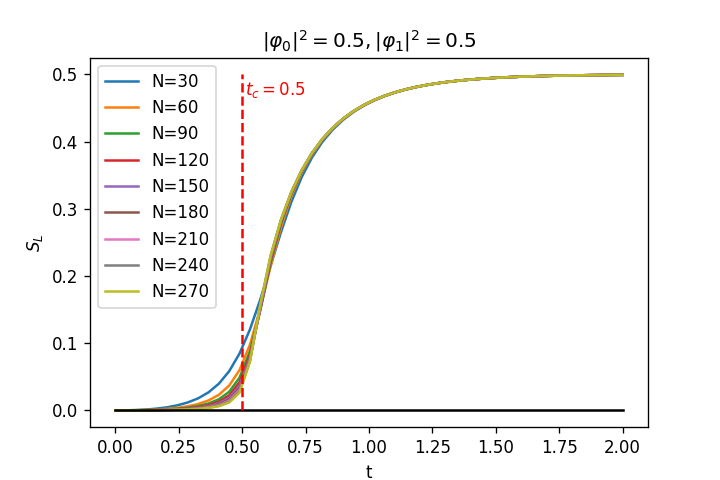}
    \caption{Initial conditions with $|\varphi_0|^2=|\varphi_1|^2=\frac{1}{2}$. Time is plotted in the horizontal axis. For $t>\frac{1}{2}$, the linear entropy has a nonzero limit for $N\to\infty$.}
    \label{fig:Impurity_t_at_05}
\end{subfigure}

\begin{subfigure}[t]{0.49\textwidth}
    \includegraphics[height=5.5cm]{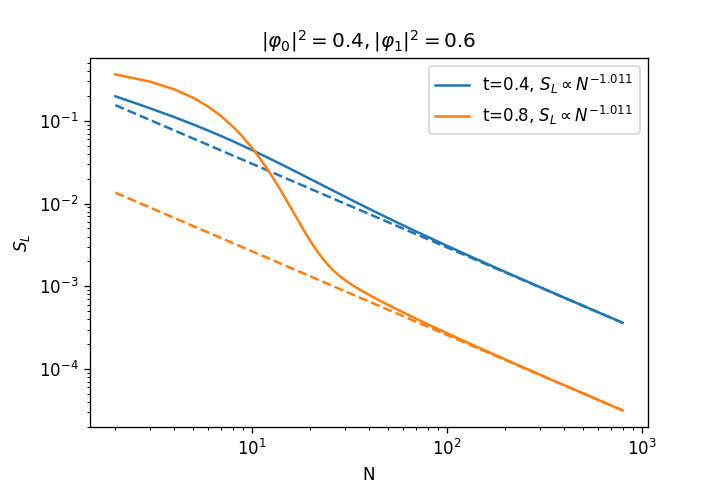}
    \caption{$N$ is plotted in the horizontal axis. The linear entropy scales as $\mathcal{O}(N^{-1})$ at every time.}
    \label{fig:Impurity_N_at_04}
\end{subfigure}
\hfill
\begin{subfigure}[t]{0.49\textwidth}
    \includegraphics[height=5.5cm]{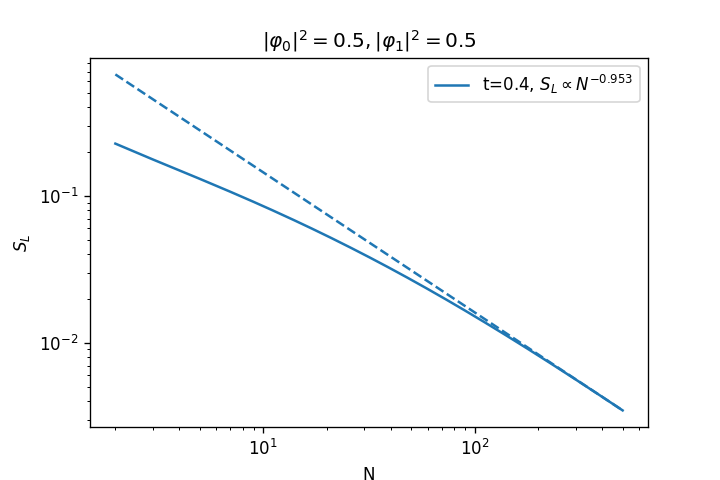}
    \caption{$N$ in plotted in the horizontal axis. The linear entropy scales as $\mathcal{O}(N^{-1})$ for $t<\frac{1}{2}$.}
    \label{fig:Impurity_N_at_05}
\end{subfigure}
\caption{{\bf Purely anti-Hermitian Hamiltonian:} Linear entropy of the one-particle reduced marginal state of the solution of the $N$-particle equation \eqref{eq:rho_entries}. Solid lines show the numerical computation of the linear entropy. We fit the tails of the curves (last 20\%) with the function $a N^{b}$. The fit is plotted in dashed lines and the values of the exponents $b$ are written in the legends.}
\label{fig:Impurity}
\end{figure}

In \autoref{fig:Impurity}, we plot the linear entropy \eqref{def:Impurity} of the one-particle marginal state \eqref{eq:rho_entries}:
\begin{itemize}
    \item In \autoref{fig:Impurity_t_at_04}, we show that for initial conditions such that $|\varphi_0|^2\neq|\varphi_1|^2$, the linear entropy tends to zero for $N\to\infty$.
    \item In \autoref{fig:Impurity_t_at_05}, we show that for initial conditions such that $|\varphi_0|^2=|\varphi_1|^2=\frac{1}{2}$, for $N\to\infty$ the linear entropy tends to zero up to $t=\frac{1}{2}$, but has a nonzero limit for $t>\frac{1}{2}$.
    \item In \autoref{fig:Impurity_N_at_04}, we show that the linear entropy scales as $\mathcal{O}(N^{-1})$ for $N\to\infty$ for initial conditions such that $|\varphi_0|^2\neq|\varphi_1|^2$.
    \item In \autoref{fig:Impurity_N_at_05}, we show that the linear entropy still scales as $\mathcal{O}(N^{-1})$ for $N\to\infty$ for initial conditions such that $|\varphi_0|^2=|\varphi_1|^2=\frac{1}{2}$ for $t<\frac{1}{2}$.
\end{itemize}

\subsubsection{Infidelity with respect to the solution of the Hartree equation}
\begin{figure}[h!t]
\centering
\begin{subfigure}[t]{0.49\textwidth}
    \includegraphics[height=5.5cm]{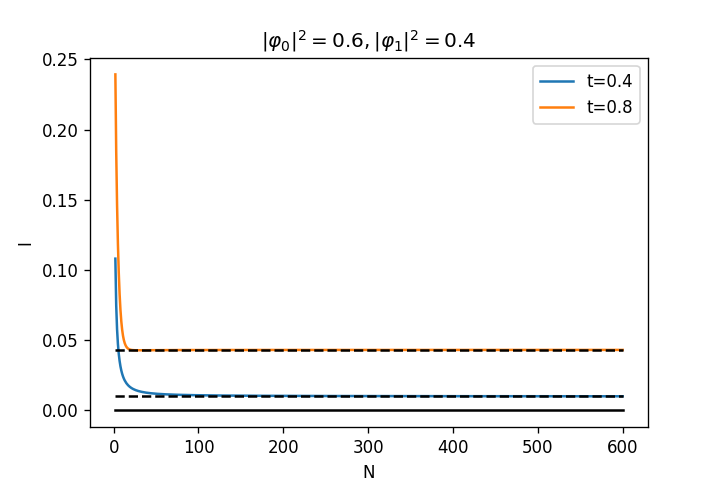}
    \caption{For initial conditions such that $|\varphi_0|^2\neq|\varphi_1|^2$, the infidelity has a nonzero limit for $N\to\infty$, and the Hartree equation does not describe the time evolution of the one-particle marginal state.}
    \label{fig:Infidelity_N_at_04}
\end{subfigure}
\hfill
\begin{subfigure}[t]{0.49\textwidth}
    \includegraphics[height=5.5cm]{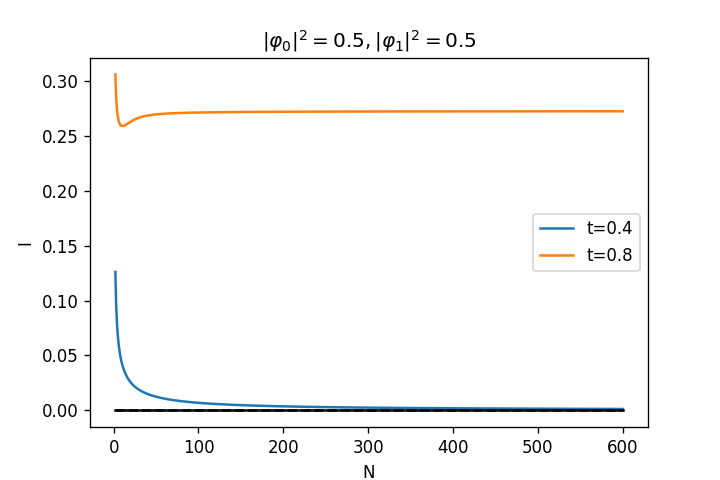}
    \caption{For initial conditions such that $|\varphi_0|^2=|\varphi_1|^2=\frac{1}{2}$, the infidelity tends to zero for $N\to\infty$ for $t<\frac{1}{2}$, but has a nonzero limit for $t>\frac{1}{2}$. Therefore, the Hartree equation describes the time evolution of the one-particle marginal state only up to $t=\frac{1}{2}$.}
    \label{fig:Infidelity_N_at_05}
\end{subfigure}

\caption{{\bf Purely anti-Hermitian Hamiltonian:} Numerical computation of the infidelity of the one-particle marginal state \eqref{eq:rho_entries} with respect to the solution of the Hartree equation \eqref{eq:HF_solution}.
Dashed lines corresponds to the limit of the infidelity for $N\to\infty$ using the results of \autoref{sec:single_maxima}.}
\label{fig:Iinfidelity}
\end{figure}

In \autoref{fig:Iinfidelity}, we analyze the infidelity \eqref{def:Infidelity} of the one-particle marginal state with respect to the solution of the Hartree equation:
\begin{itemize}
\item In \autoref{fig:Infidelity_N_at_04}, we show that for initial conditions such that $|\varphi_0|^2\neq|\varphi_1|^2$, the infidelity has a nonzero limit for $N\to\infty$, and therefore the Hartree equation does not describe the time evolution of the one-particle marginal state.
\item In \autoref{fig:Infidelity_N_at_05}, we show that for initial conditions such that $|\varphi_0|^2=|\varphi_1|^2=\frac{1}{2}$, the infidelity tends to zero for $N\to\infty$ up to $t=\frac{1}{2}$, but has a nonzero limit for $t>\frac{1}{2}$. Therefore, the Hartree equation correctly describes the time evolution of the one-particle marginal state only up to $t=\frac{1}{2}$.
\end{itemize}

\subsubsection{Convergence speed towards the limit for \texorpdfstring{$N\to\infty$}{N→∞}}

The last numerical analysis concerns the speed of the convergence of the one-particle marginal state towards its limit for $N\to\infty$.
We define the infidelity of the one-particle marginal state $\hat{\rho}^{(1)}$ with respect to its limit for $N\to\infty$ as
\begin{equation} \label{eq:infidelity_largeN}
    \varepsilon = 1 -F\left(\hat{\rho}^{(1)},\lim_{N\to\infty} \hat{\rho}^{(1)}\right)\,,
\end{equation}
where the fidelity between the quantum states $\rho$ and $\sigma$ is
\begin{equation}
    F(\rho,\sigma) = \left( \operatorname{Tr} \sqrt{\sqrt{\rho} \, \sigma \sqrt{\rho}} \right)^2
\end{equation}
and the value of $x^*_t$ is found by numerically minimizing $f_t(x)$.

\begin{figure}[h!t]
\centering
\begin{subfigure}[t]{0.49\textwidth}
    \includegraphics[height=5.5cm]{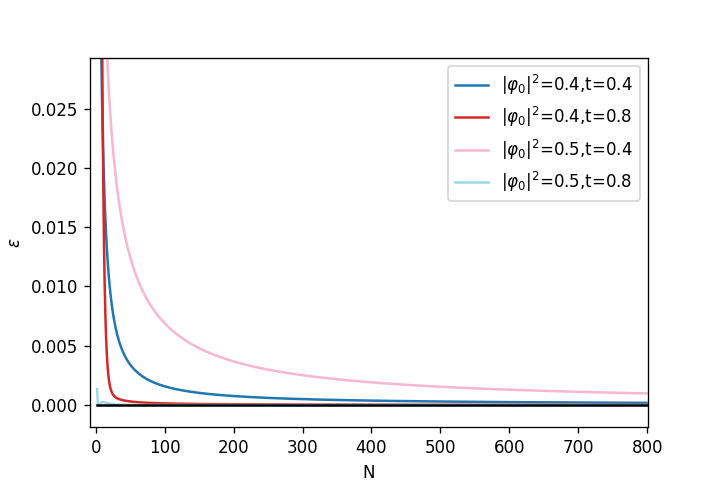}
    \caption{Linear-scale plot of the infidelity with respect to the limit $N\to\infty$ as a function of $N$, at different instances of the initial condition and time.}
    \label{fig:LargeNlim_linear}
\end{subfigure}
\hfill
\begin{subfigure}[t]{0.49\textwidth}
    \includegraphics[height=5.5cm]{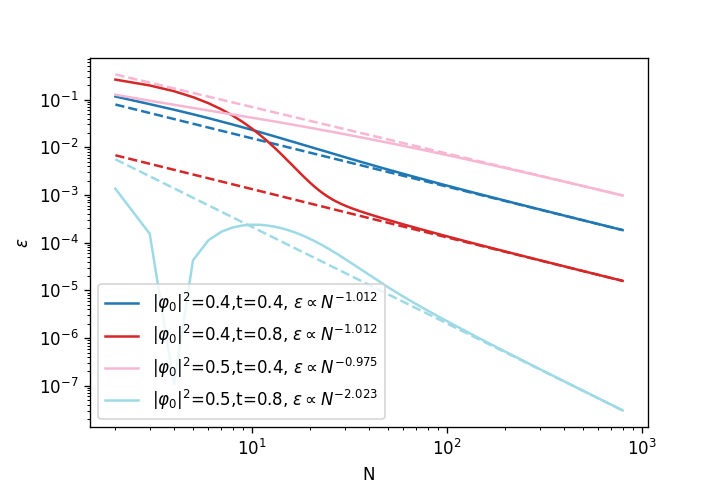}
    \caption{We realized a fit to the tails (last 20\%) of the curves with a function $aN^b$. The results are plotted with dashed lines. The obtained exponents $b$ are written in the legend.}
    \label{fig:LargeNlim_log}
\end{subfigure}

\caption{{\bf Purely anti-Hermitian Hamiltonian:} Numerical computation of the infidelity of the one-particle marginal state \eqref{eq:rho_entries} with respect to its limit \eqref{eq:lim_N_rho(1)} for $N\to\infty$. The infidelity scales as $\mathcal{O}(N^{-1})$ for $N\to\infty$ for almost all initial conditions, and scales as $\mathcal{O}(N^{-2})$ for the initial conditions such that $|\varphi_0|^2=|\varphi_1|^2=\frac{1}{2}$ and $t>\frac{1}{2}$.}
\label{fig:LargeNlim}
\end{figure}

In \autoref{fig:LargeNlim_linear}, we show that the infidelity always tends to zero for $N\to\infty$ (as it should).
In \autoref{fig:LargeNlim_log}, we show that the infidelity scales as $\mathcal{O}(N^{-1})$ for $N\to\infty$ both for initial conditions such that $|\varphi_0|^2\neq|\varphi_1|^2$ and for initial conditions such that $|\varphi_0|^2 = |\varphi_1|^2 = \frac{1}{2}$ and $t<\frac{1}{2}$, matching the scaling in the Hermitian case.
Instead, for initial conditions such that $|\varphi_0|^2=|\varphi_1|^2=\frac{1}{2}$ and $t>\frac{1}{2}$, the infidelity scales as $\mathcal{O}(N^{-2})$.

\section{Second counterexample: Hamiltonian with real spectrum}\label{sec:ZZ_RealSpectrum}

As a counterexample to the mean-field approximation, we considered in \autoref{sec:ZZ_hamiltonian} a purely anti-Hermitian Hamiltonian. Since its eigenvalues are purely imaginary, the dynamics of the normalized wavefunction converges exponentially fast to the eigenvectors with maximum eigenvalues. The same effect would take place for a generic purely anti-Hermitian Hamiltonian: the wavefunction would converge exponentially fast to the eigenvector associated to the eigenvalue with largest imaginary part. In order to avoid this effect, we construct an example of a non-Hermitian Hamiltonian with real spectrum.

As in the previous example, we take $\mathfrak{h}=\mathbb{C}^2$, the qubit Hilbert space,  choose a non-orthogonal basis of $\mathfrak{h}$ given by normalized vectors $\{|v_0\rangle, |v_1\rangle \}$, and define the change of basis given by the operator
\begin{equation}
    C := |v_0\rangle \langle0| + |v_1\rangle \langle 1 |. 
\end{equation}
Let us consider the interaction between two particles  given by $A^{(2)}=CZC^{-1} \otimes CZC^{-1}$.
The corresponding mean-field Hamiltonian for $N$ particles is given by
\begin{equation}\label{def:ZZ_realspectrum}
    A^{(N)} = \frac{1}{N-1} \sum_{1\leq i<j\leq N} (CZC^{-1})_i (CZC^{-1})_j.
\end{equation}
In order to diagonalize it, we rewrite it as 
\begin{equation}
    A^{(N)} = C^{\otimes N} \frac{1}{2(N-1)} \left( \left(\sum_{i=1}^{N}Z_i\right)^2 - N \right) \left(C^{-1}\right)^{\otimes N}.
\end{equation}
Similarly to \eqref{eq:eigenvector}, the eigenvalues and eigenvectors of $A^{(N)}$ are given by
\begin{equation}
    A^{(N)}C^{\otimes N} \ket{N,n} = \lambda_{N,n}C^{\otimes N} \ket{N,n} , \quad \lambda_{N,n} = \frac{(N-2n)^2-N}{2(N-1)}.
\end{equation}
Observe that the spectrum is real. We quantify the non-Hermiticity induced by the change of basis $C$ with the overlap $\alpha = \langle v_0|v_1\rangle$. Then we can compute
\begin{equation}
    C^\dagger C =  \mathbb{I} + \alpha |v_0\rangle \langle v_1| + \alpha^* |v_1\rangle \langle v_0 |.
\end{equation}
Even if the spectrum is real, since the Hamiltonian is not Hermitian, the eigenvectors of $A^{(N)}$ are not orthogonal, instead they satisfy

\begin{equation}\label{eq:nonOrthogonality}
    \langle N,n| C^{\dagger \otimes N} C^{\otimes N} |N,m\rangle = \frac{1}{\sqrt{\binom{N}{m}\binom{N}{n}}} \sum_a \alpha^{n-a} \alpha^{* m-a} \binom{N}{a,n-a,m-a,N-m-n+a}.
\end{equation}

Before solving the Schr\"odinger equation \eqref{eq:Many-body_Schrodinger_Psi} with initial state $|\Psi^{(N)}(0)\rangle = |\varphi\rangle ^{\otimes N}$, we need to write the initial state in the eigenbasis. Let us define $|\eta \rangle=\eta_0|0\rangle+\eta_1|1\rangle := C^{-1}|\varphi\rangle$, which is not normalized if $C$ is not unitary. So the initial state is given by $|\Psi^{(N)}(0)\rangle = (C |\eta\rangle)^{\otimes N}$. Then the decomposition of the initial state in the eigenbasis of the non-Hermitian Hamiltonian is given by

\begin{equation}
    |\Psi^{(N)}(0)\rangle = \sum_{n=0}^{N} \eta_0^n \eta_1^{N-n} \sqrt{\binom{N}{n}} C^{\otimes N} \ket{N,n},
\end{equation}
the time evolution is given by
\begin{equation}
    |\Psi^{(N)}(t)\rangle = \sum_{n=0}^{N} \eta_0^n \eta_1^{N-n} \sqrt{\binom{N}{n}} e^{-i\lambda_{N,n}t} C^{\otimes N} \ket{N,n},
\end{equation}
and the one-particle reduced density matrix is given by 
\begin{equation}
    \rho^{(1)} =  \sum_{m,n=0}^{N} \eta_0^n \eta_1^{N-n}\eta_0^{*m} \eta_1^{*N-m} \sqrt{\binom{N}{n}\binom{N}{m}} e^{-i(\lambda_{N,n}-\lambda_{N,m})t}\ \Tr_{2\dots N}\left( C^{\otimes N} |N,n\rangle\langle N,m|C^{\dagger\otimes N}\right)
\end{equation}

Using the non-orthogonality \eqref{eq:nonOrthogonality}, we can write the exact  evolution of the reduced density matrix in the chosen basis: 
\begin{equation} \label{eq:rho(1)_ZZ_real}
\begin{split}
     \rho^{(1)} = & \sum_{\underset{\sum n_l=N-1}{n_0,n_1,n_2,n_3 \geq0}} |\eta_0|^{2n_0}|\eta_1|^{2n_1}(\eta_0\eta^*_1\alpha^*)^{n_2}(\eta^*_0\eta_1\alpha)^{n_3} \binom{N-1}{n_0,n_1,n_2,n_3} \\& \times\left( \sum_{i,j=0}^{1} e^{-it\left(\lambda_{N,n_2+n_0+1-i} -\lambda_{N,n_3+n_0+1-j}\right)} \eta_i|v_i\rangle\langle v_j| \eta^*_j \right).
\end{split}
\end{equation}

We are going to compare the exact normalized single particle evolution with the non-Hermitian Hartree equation \eqref{eq:HF_non_Hermitian_vector}. In this example, it takes the form 

\begin{equation}\label{eq:HF_ZZ_real}
    i \partial_t |\phi\rangle = \langle\phi|CZC^{-1}|\phi\rangle CZC^{-1}|\phi\rangle  - i\ \text{Im}\left(\langle \phi|CZC^{-1} |\phi\rangle^2\right) |\phi\rangle. 
\end{equation}

\subsection{Hermitian case}

We start by analyzing the Hermitian case in order to understand how the reduced density matrix actually converges to the solution of the Hartree equation. Set $\alpha=0$ in \eqref{eq:rho(1)_ZZ_real} to compute the reduced density matrix. Using that $\alpha^n = \delta_{n,0}$, and the binomial theorem, the reduced density matrix simplifies to: 
\begin{equation}
\begin{split}
    \rho^{(1)} =& |\eta_0|^2 |v_0\rangle\langle v_0| + \eta_0\eta_1^*\left( |\eta_0|^2 e^{-i\frac{2t}{N-1}} + |\eta_1|^2 e^{i\frac{2t}{N-1}}\right)^{N-1} |v_0\rangle\langle v_1|\\
    +& \eta_0^*\eta_1\left( |\eta_0|^2 e^{i\frac{2t}{N-1}} + |\eta_1|^2 e^{-i\frac{2t}{N-1}}\right)^{N-1} |v_1\rangle\langle v_0| + |\eta_1|^2 |v_1\rangle\langle v_1|.
\end{split}
\end{equation}
A straightforward calculation of the linear entropy, defined in \eqref{def:Impurity}, gives
\begin{equation}\label{eq:S_L_ZZ_Hermitian}
    S_L(\rho^{(1)})= 2 |\eta_0|^2|\eta_1|^2\left( 1-\left[1+2|\eta_0|^2|\eta_1|^2\left(\cos\left(\frac{4t}{N-1}\right)-1\right)\right]^{(N-1)}\right).
\end{equation}
We argued that the mean-field approximation requires that $\lim_{N\to \infty}S_L(\rho^{(1)})= 0$ at any fixed time and indeed, it is true for \eqref{eq:S_L_ZZ_Hermitian}. If we expand it for large $N$ we obtain

\begin{equation}\label{eq:S_L_ZZ_Hermitian_LargeN}
    S_L(\rho^{(1)}) = \frac{(8|\eta_0|^2|\eta_1|^2t)^2}{N}+\mathcal{O}\left(\frac{1}{N^2}\right).
\end{equation}
But we make the observation that for any fixed number of particles, the linear entropy is a periodic function of time with period $(N-1)/8\pi$ and it has the maximum value 
\begin{equation}
    \max_{t}S_L(\rho^{(1)}) = 2|\eta_0|^2|\eta_1|^2\left( 1- \left(1-4|\eta_0|^2|\eta_1|^2\right)^{N-1}\right).
\end{equation}
This value increases with $N$ and tends to $2|\eta_0|^2|\eta_1|^2$ for $N\to\infty$. So the maximum possible value of the linear entropy over time increases with the number of particles, but the time required to achieve this value increases linearly in the number of particles, in such a way that when time is fixed we have the convergence towards a pure state following \eqref{eq:S_L_ZZ_Hermitian_LargeN}.

Since the one-particle reduced density matrix converges to a pure state, we can compare it with the solution of the Hartree equation \eqref{eq:HF_ZZ_real}. Since $\alpha=0$, then $C^{-1}=C^\dagger$, then the Hartree equation  takes the form
\begin{equation}\label{eq:HF_ZZ_hermitian}
    i \partial_t |\phi\rangle = \langle\phi,CZC^{\dagger}\phi\rangle C ZC^{\dagger} |\phi\rangle,
\end{equation}
with initial value $|\phi(0)\rangle=C|\eta\rangle$. Given that $z_0 = \langle\phi,CZC^{\dagger
}\phi\rangle$ is a constant of motion, the solution to \eqref{eq:HF_ZZ_hermitian} is $|\phi\rangle = Ce^{-iz_0 Z t}|\eta\rangle$. We can then compute the infidelity \eqref{def:Infidelity} between the exact one-particle reduced density matrix and the Hartree approximation and we get
\begin{equation}
    I = 2|\eta_0|^2|\eta_1|^2\left(1- \text{Re}\left( \left( |\eta_0|^2 e^{i\frac{1-z_0}{N-1}t}+ |\eta_1|^2 e^{-i\frac{1+z_0}{N-1}t}\right)^{N-1} \right)\right).
\end{equation}
Similarly to the linear entropy \eqref{eq:S_L_ZZ_Hermitian}, the infidelity is also a periodic function of time, with period that grows linearly with the number of particles and at fixed time it converges to zero for $N\to\infty$:
\begin{equation}
    I = \frac{16|\eta_0|^4|\eta_1|^4t^2}{N} + \mathcal{O}\left(\frac{1}{N^2}\right).
\end{equation}
This shows, as it is known for all Hermitian Hamiltonians, that the Hartree equation correctly reproduces the one-particle reduced density matrix for any fixed time in the limit $N\to\infty$.

\subsection{Numerical analysis of the non-Hermitian case}

When $\alpha\neq0$, the Hartree equation \eqref{eq:HF_ZZ_real} is not solvable analytically, and the sum in \eqref{eq:rho(1)_ZZ_real} cannot be simplified using the binomial theorem. So we proceed with the numerical computation of these quantities. 

The Hartree equation can easily be integrated numerically with standard methods, in particular we use the explicit Runge-Kutta method of order 5 implemented in the library Scipy in Python.
The numerical computation of \eqref{eq:rho(1)_ZZ_real} is not stable since there might be very large numbers coming from the multinational coefficient $\binom{N-1}{n_0,n_1,n_2,n_3}=\tilde{\mathcal{O}}(4^N)$, with different phases, coming from $(\eta_0\eta^*_1\alpha^*)^{n_2}(\eta^*_0\eta_1\alpha)^{n_3}$. Computing such differences between large numbers requires a float precision that scales linearly with $N$. This approach is very expensive computationally and it does not allow to reach large values of $N$. In order to avoid this large cancellations problem, we restrict the initial data such that $\text{Re}(\alpha \eta_0^*\eta_1)>0$. We observed numerically that in these cases the numerical instabilities were not a problem so we can compute $\rho^{(1)}$ using 64-bit floats.

\begin{figure}[h!t]
\centering
\begin{subfigure}[t]{0.49\textwidth}
    \includegraphics[width=\textwidth]{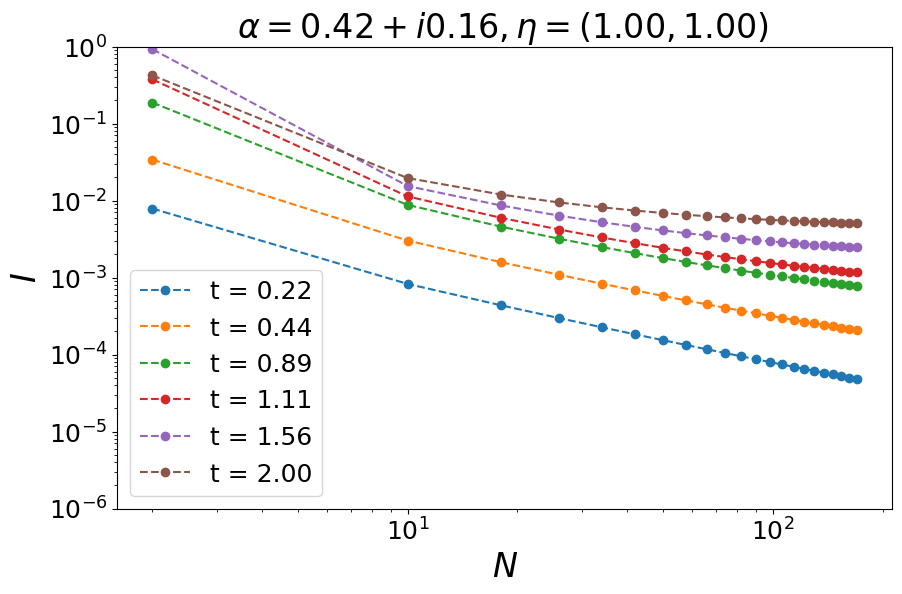}
    \caption{The infidelity converges to a non-zero value.}
    \label{fig:infidelity-vs_N_alpha_inC_balanced}
\end{subfigure}
\hfill
\begin{subfigure}[t]{0.49\textwidth}
    \includegraphics[width=\textwidth]{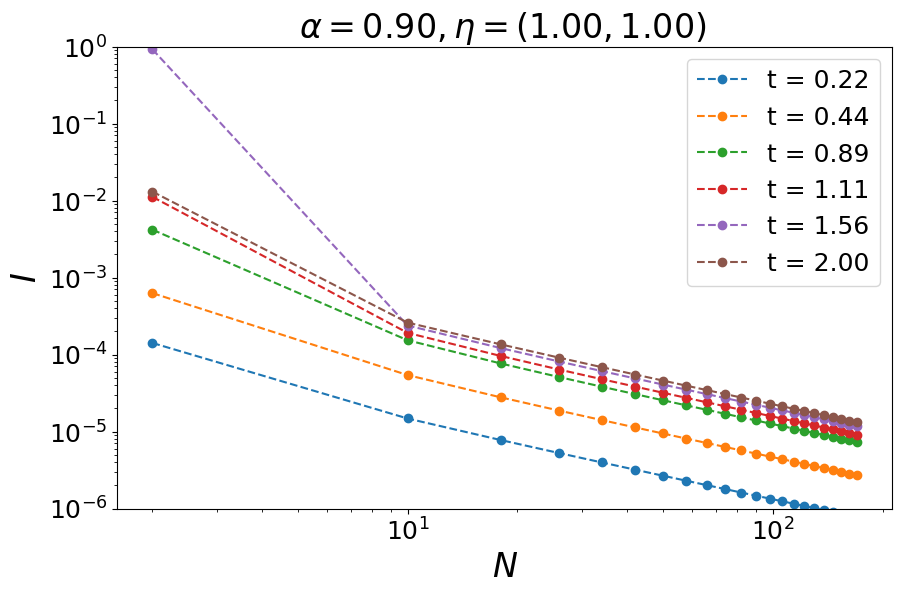}
    \caption{The infidelity vanishes at large $N$.}
    \label{fig:Infidelity-vs-N_alpha_inR_eta_balanced}
\end{subfigure}

\begin{subfigure}[t]{0.49\textwidth}
    \includegraphics[width=\textwidth]{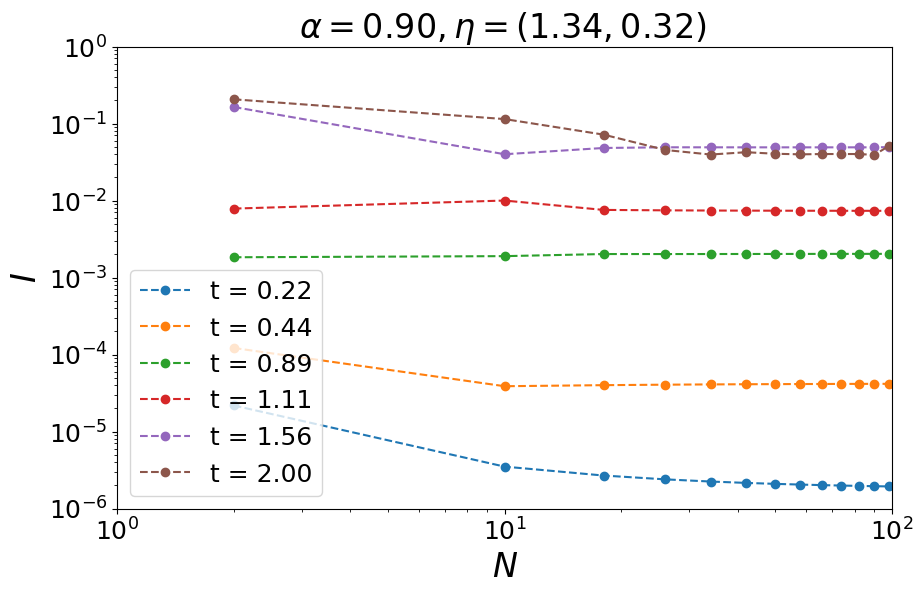}
    \caption{The infidelity converges very quickly to a non-zero value.}
    \label{fig:Infidelity-vs-N_alpha_inR_eta_notBalanced}
\end{subfigure}
\caption{{\bf Non-Hermitian Hamiltonian with real spectrum:} We numerically compute the infidelity \eqref{def:Infidelity} between the reduced density matrix \eqref{eq:rho(1)_ZZ_real} and the solution of the Hartree equation \eqref{eq:HF_ZZ_real}. We plot $I$ as a function of the number of particles $N$ in log-log scale for fixed times, and for three different choices of $\alpha$ and $\eta$. Observe that \autoref{fig:infidelity-vs_N_alpha_inC_balanced} and \autoref{fig:Infidelity-vs-N_alpha_inR_eta_balanced} have the same $\eta$, and \autoref{fig:Infidelity-vs-N_alpha_inR_eta_balanced} and \autoref{fig:Infidelity-vs-N_alpha_inR_eta_notBalanced} have the same $\alpha$. The infidelity converges to zero only in \autoref{fig:Infidelity-vs-N_alpha_inR_eta_balanced}.} 
\label{fig:Infidelity_ZZ_Real}
\end{figure}

By computing several instances satisfying $\text{Re}(\alpha \eta_0^*\eta_1)>0$, we observe that the infidelity does not always vanish for large $N$, showing that even if the eigenvalues of the Hamiltonian are real, non-Hermiticity can break the mean-field approximation. To depict the possible behaviors of the system as a function of the number of particles, we plot in \autoref{fig:Infidelity_ZZ_Real} the infidelity defined in \eqref{def:Infidelity} for three different instances.

In \autoref{fig:Infidelity-vs-N_alpha_inR_eta_balanced}, where $\alpha>0$ and $\eta_0=\eta_1$, the infidelity vanishes, showing that the approximation holds for this particular case. However, if we change either $\alpha$ (\autoref{fig:infidelity-vs_N_alpha_inC_balanced}) or $\eta$ (\autoref{fig:Infidelity-vs-N_alpha_inR_eta_notBalanced}), the approximation breaks and the infidelity converges to some nonzero value.

\subsection{Second time derivative}

We are going to use the equation \eqref{eq:diff_secondderivatives} to analyze the set of initial conditions $\eta$ and non-Hermiticity $\alpha$ such that the Hartree equation describes correctly the second time derivative. In order to simplify the notation we define $\mathcal{Z}=CZC^{-1}$. In this example, equation \eqref{eq:diff_secondderivatives} reads

\begin{equation}
\begin{split}
    \partial_t^2(&\hat{\rho}^{(1)}-\gamma)|_{t=0}=\\
    &\left[-\langle\varphi| \mathcal{Z}|\varphi\rangle +\langle\varphi|\mathcal{Z}^\dagger\mathcal{Z} |\varphi\rangle +\langle\varphi| \mathcal{Z}|\varphi\rangle^3 - \langle\varphi|\mathcal{Z}^\dagger |\varphi\rangle^2\langle\varphi|\mathcal{Z} |\varphi\rangle \right]\left(\mathcal{Z}-\langle\varphi|\mathcal{Z} |\varphi\rangle\right)  |\varphi\rangle\langle\varphi| + h.c.\\
    &+ \mathcal{O}(N^{-1}).
\end{split}
\end{equation}
A necessary condition for the Hartree approximation to describe the system is that this equation must vanish at the first order. This happens if one of the two terms in parenthesis vanishes, lets analyze each case separately.

The first case is given by the condition
\begin{equation}\label{eq:first_cond}
    \left(\mathcal{Z}-\langle\varphi|\mathcal{Z} |\varphi\rangle\right)  |\varphi\rangle = 0.
\end{equation}
This equation implies that the initial condition $\varphi$ is an eigenvector of the operator $\mathcal{Z}$. In this case, the evolution of the Hartree solution and of the many-body equation  is trivial and then the Hartree approximation is exact.

The second case is given by the condition 
\begin{equation}\label{eq:second_cond}
    -\langle\varphi| \mathcal{Z}|\varphi\rangle +\langle\varphi|\mathcal{Z}^\dagger\mathcal{Z} |\varphi\rangle +\langle\varphi| \mathcal{Z}|\varphi\rangle^3 - \langle\varphi|\mathcal{Z}^\dagger |\varphi\rangle^2\langle\varphi|\mathcal{Z} |\varphi\rangle = 0.
\end{equation}
This equation can be interpreted by observing the Hartree evolution \eqref{eq:HF_ZZ_real}. Using this  equation one can compute the time derivative of the expectation value of the operator $\mathcal{Z}$:
\begin{equation}
    i\partial_t\langle\phi| \mathcal{Z}|\phi\rangle = \langle\phi| \mathcal{Z}|\phi\rangle -\langle\phi|\mathcal{Z}^\dagger\mathcal{Z} |\phi\rangle -\langle\phi| \mathcal{Z}|\phi\rangle^3 + \langle\phi|\mathcal{Z}^\dagger |\phi\rangle^2\langle\phi|\mathcal{Z} |\varphi\rangle.
\end{equation}
Then, the condition \eqref{eq:second_cond} implies that $\partial_t\langle\phi| \mathcal{Z}|\phi\rangle|_{t=0} =0$.

We can write explicitly this condition in terms of the  variables $|\eta\rangle=C^{-1}|\varphi\rangle$ and $\alpha$:
\begin{equation}
    2\left(|\eta_0|^2-|\eta_1|^2 \right) \left(\text{Re}\left(\alpha^*\eta_0\eta_1^*\right)+\text{Im}\left(\alpha^*\eta_0\eta_1^*\right)^2\right)+ i \text{Im}\left(\alpha^*\eta_0\eta_1^*\right)\left(|\eta_0|^2+|\eta_1|^2-2\left(|\eta_0|^2-|\eta_1|^2\right)^2 \right) = 0\ .
\end{equation}
One can observe in the numerical examples that the instance simulated in \autoref{fig:Infidelity-vs-N_alpha_inR_eta_balanced} satisfies this condition and indeed the Hartree approximation works, while for the instances simulated in \autoref{fig:Infidelity-vs-N_alpha_inR_eta_notBalanced} and \autoref{fig:infidelity-vs_N_alpha_inC_balanced} this condition is not satisfied and the approximation fails.

\section{Conclusions and perspectives}\label{sec:concl}

In this paper, we have proved that the Hartree equation does not always provide a good approximation to the evolution of the one-particle marginal state under a non-Hermitian many-body dynamics.
 First, we have proved that the Hartree equation is not able to capture the right second time derivative of the one-particle marginal state at $t=0$ in the limit $N\to\infty$.
This failure can be ascribed to the anti-hermitian terms of the Hamiltonian that do not act on the considered particle but provide a contribution to the time derivative of the one-particle state that amplifies the deviations from the Hartree equation and does not vanish when taking the partial trace over the other particles nor when taking the limit $N\to\infty$. 

Then, we have considered a system of $N$ bosonic qubits evolving with the purely anti-Hermitian Hamiltonian $\frac{i}{N-1}\sum_{i<j}Z_i\,Z_j$ starting from a pure product state, and we have computed an analytic expression for the limit for $N\to\infty$ of the one-particle marginal state.
We have found that, for almost every initial condition, such a limit is pure, but does not satisfy the non-Hermitian generalization of the Hartree equation. Therefore, the Hartree approximation fails even if the limit of the one-particle marginal state is pure.
Furthermore, we have derived the correct effective evolution equation for the limiting one-particle wavefunction. This evolution differs from the Hartree equation by an explicit time dependence, and it matches the Hartree
equation only at $t=0$.
Moreover, we have shown that there exists an initial condition such that the one-particle dynamics exhibits a finite critical time $t_c=\tfrac12$. The limit for $N\to\infty$ of the one-particle marginal state is pure and satisfies the Hartree equation up to $t=\frac{1}{2}$, but for $t>\frac{1}{2}$ the limiting one-particle marginal state becomes mixed and the Hartree approximation breaks down.

Furthermore, we have numerically studied a bosonic non-Hermitian mean-field Hamiltonian with real spectrum and we have showed that also in this case, there are initial product states such that the time-evolved one-particle state does not converge to the solution of the Hartree equation in the limit $N\to\infty$.

Our findings impose strong restrictions on the validity of mean-field theory for many-body bosonic non-Hermitian Hamiltonians.

A direct motivation for understanding non-Hermitian mean-field limits comes from the quantum algorithm for
nonlinear differential equations of \cite{NonLinearDiffEq_Lloyd}, which proposes to solve a
nonlinear evolution by embedding it into a linear evolution on many copies and then invoking a Hartree/mean-field
reduction on the copies. Our counterexample shows that the validity of this
reduction cannot be assumed generically: even in an exactly solvable model with two-body interactions, the limit for $N\to\infty$ of the one-particle marginal states may not satisfy the Hartree equation. Consequently, the nonlinear dynamics obtained from a many-copy linear embedding can differ from the intended target equation unless additional conditions are imposed. From the algorithmic perspective, this suggests that one must either restrict to classes of nonlinear differential equations for which the validity of the Hartree approximation can be proved or incorporate a certification/verification step that confirms that the reduced dynamics follows the desired effective nonlinearity for the specific instance at hand.

Non-Hermitian Hamiltonians are routinely used in many-body physics to model particle gain and loss, both as phenomenological effective theories and as conditional no-jump generators in quantum-trajectory descriptions of Lindblad dynamics. Our results show that non-unitarity can qualitatively alter the mean-field mechanism: even when the limit of the one-body marginal state is pure, the correct effective dynamics may acquire additional (in general, explicitly time-dependent) structure not captured by the Hartree approximation, and multiple competing
macroscopic contributions can produce mixed limiting marginals after a finite time. This indicates that
care is needed when using non-Hermitian Hartree equations to predict one-body observables in lossy
many-body settings.

Our counterexamples motivate two complementary research directions. The first one is to characterize the set of non-Hermitian Hamiltonians for which the Hartree approximation is valid. We already know that such set includes non-interacting systems and arbitrary bosonic systems with Hermitian Hamiltonians.
A hint for such a set can be given by the non-Hermitian Hamiltonians such that at $t=0$, the second time derivative of the one-particle marginal state matches the second time derivative of the solution of the Hartree equation.

The second possible direction is more challenging and consists in deriving a correct effective evolution equation for the one-particle marginal state. In this paper, we have derived such an equation for the considered Hamiltonian \eqref{eq:lim_dtdnu}.  Since this equation includes an explicit time dependence despite the many-body Hamiltonian is time-independent, our findings suggest that a general framework to describe non-Hermitian dynamics of bosonic mean-field systems may need to include additional information not captured by the naive product-state ansatz. Establishing such a framework and clarifying its relationship to mean-field limits would significantly improve the reliability of mean-field models in both many-body quantum physics and quantum algorithm design.

\section*{Acknowledgements}
GDP has been supported by the HPC Italian National Centre for HPC, Big Data and Quantum Computing -- Proposal code CN00000013 -- CUP J33C22001170001 and by the Italian Extended Partnership PE01 -- FAIR Future Artificial Intelligence Research -- Proposal code PE00000013 -- CUP J33C22002830006 under the MUR National Recovery and Resilience Plan funded by the European Union -- NextGenerationEU.
Funded by the European Union -- NextGenerationEU under the National Recovery and Resilience Plan (PNRR) -- Mission 4 Education and research -- Component 2 From research to business -- Investment 1.1 Notice Prin 2022 -- DD N. 104 del 2/2/2022, from title ``understanding the LEarning process of QUantum Neural networks (LeQun)'', proposal code 2022WHZ5XH -- CUP J53D23003890006.
GDP and DP are members of the ``Gruppo Nazionale per la Fisica Matematica (GNFM)'' of the ``Istituto Nazionale di Alta Matematica ``Francesco Severi'' (INdAM)''. SR is supported by the European Research
Council via the ERC CoG RAMBAS–Project–Nr. 10104424.

\section*{Author Contributions}
    All the authors contributed equally to this work.

\printbibliography

\end{document}